\documentclass[a4paper, 12pt, numbers = noenddot, parskip = no]{scrartcl}
\usepackage[top = 2cm, bottom = 4cm]{geometry}

\usepackage{authblk}
\usepackage[T1]{fontenc}
\usepackage[utf8]{inputenc}
\usepackage{lmodern}
\usepackage{siunitx}

\setkomafont{sectioning}{\normalfont \rmfamily \bfseries}%
\setkomafont{paragraph}{\normalfont \rmfamily \bfseries}%

\usepackage{graphicx}
\graphicspath{{./fig/}}
\usepackage{subcaption}

\usepackage{mathtools} 
\mathtoolsset{showonlyrefs=true}
\usepackage[ntheorem, framemethod=TikZ]{mdframed} 
\usepackage{framed} 
\usepackage[amsmath, hyperref, framed]{ntheorem}%
\usepackage{amssymb} 
\usepackage{bm}

\DeclareMathOperator{\dBetaBin}{BetaBinom} 
\DeclareMathOperator{\dUnif}{Unif} 
\DeclareMathOperator{\dBin}{Binom} 
\DeclareMathOperator{\dBeta}{Beta} 
\DeclareMathOperator{\dNegBin}{Neg-Bin} 

\newcommand{\intDiff}{\,\mathrm{d}} 
\DeclareMathOperator{\E}{\mathbb{E}} 
\DeclareMathOperator{\var}{var} 
\DeclareMathOperator{\ind}{\mathbf{1}} 
\newcommand{\ccdot}{\,\cdot\,} 
\DeclareMathOperator{\logit}{logit} 
\newcommand{\T}{\top} 
\DeclareMathOperator{\bo}{\mathcal{O}}%
\newcommand{\Mod}{\,\mathrm{mod}\,}

\newcommand{\calH}{\mathcal{H}}%
\newcommand{\calS}{\mathcal{S}}%
\newcommand{\calX}{\mathcal{X}}%
\newcommand{\mathsc}[1]{{\normalfont\textsc{#1}}}%

\newcommand{\dCas}{\smash{\tilde{d}}}%
\newcommand{\dCon}{\smash{\bar{d}}}%
\newcommand{\rCas}{\smash{\tilde{r}}}%
\newcommand{\rCon}{\smash{\bar{r}}}%
\newcommand{\nCas}{\smash{\tilde{n}}}%
\newcommand{\nCon}{\smash{\bar{n}}}%
\newcommand{\yCas}{\smash{\tilde{y}}}%
\newcommand{\yCon}{\smash{\bar{y}}}%
\newcommand{\xCas}{\smash{\tilde{x}}}%
\newcommand{\xCon}{\smash{\bar{x}}}%
\newcommand{\piCas}{\smash{\tilde{\pi}}}%
\newcommand{\piCon}{\smash{\bar{\pi}}}%
\newcommand{\SCas}{\smash{\tilde{S}}}%
\newcommand{\SCon}{\smash{\bar{S}}}%
\newcommand{\PCas}{\smash{\tilde{P}}}%
\newcommand{\PCon}{\smash{\bar{P}}}%
\newcommand{\nuCas}{\smash{\tilde{\nu}}}%
\newcommand{\nuCon}{\smash{\bar{\nu}}}%
\newcommand{\fCas}{\smash{\tilde{f}}}%
\newcommand{\fCon}{\smash{\bar{f}}}%
\newcommand{\gCas}{\smash{\tilde{g}}}%
\newcommand{\gCon}{\smash{\bar{g}}}%
\newcommand{\rhoCas}{\smash{\tilde{\rho}}}%
\newcommand{\rhoCon}{\smash{\bar{\rho}}}%
\newcommand{\uCas}{\smash{\tilde{u}}}%

\newcommand{\piAll}{\pi}%
\newcommand{\yAll}{y}%
\newcommand{\nAll}{n}%
\newcommand{\xAll}{x}%
\newcommand{\gAll}{g}%
\newcommand{\uAll}{u}%

\newcommand{\probSplit}{q_{\mathsc{split}}}%
\newcommand{\probMerge}{q_{\mathsc{merge}}}%

\newcommand{\hCas}{\smash{\tilde{h}}}
\newcommand{\omegaCas}{\smash{\tilde{\omega}}}
\newcommand{\kappaCas}{\smash{\tilde{\kappa}}}

\newcommand{\reals}{\mathbb{R}}%
\newcommand{\naturals}{\mathbb{N}}%
\DeclareFontFamily{U}{mathx}{\hyphenchar\font45}
\DeclareFontShape{U}{mathx}{m}{n}{<-> mathx10}{}
\DeclareSymbolFont{mathx}{U}{mathx}{m}{n}
\DeclareMathAccent{\widebar}{0}{mathx}{"73}

\usepackage[inline]{enumitem}
\newlist{myenumerate}{enumerate}{3} 
\setlist[myenumerate]{itemsep=1ex, topsep=0.7ex}
\setlist[myenumerate,1]{label=\arabic*., ref=\arabic*, leftmargin=*, itemsep=-0.1ex, topsep=0.7ex}
\setlist[myenumerate,2]{label=\roman*., ref=\roman*, labelindent=\parindent, topsep=-0.25ex, itemsep=-0.1ex}
\setlist[myenumerate,3]{label=\alph*., ref=\alph*, labelindent=\parindent, topsep=-0.25ex, itemsep=-0.1ex}

\newlist{myitemize}{enumerate}{2} 
\setlist[myitemize]{itemsep=1ex, topsep=0.5ex}
\setlist[myitemize,1]{label=\textbullet, leftmargin=*, itemsep=-0.1ex, topsep=0.6ex}
\setlist[myitemize,2]{label=$-$, labelindent=\parindent, topsep=-0.25ex, itemsep=-0.1ex}

\makeatletter
\newcommand{\mylabel}[2]{#2\def\@currentlabel{#2}\label{#1}}
\makeatother

\usepackage{lineno}

\newcommand*\patchAmsMathEnvironmentForLineno[1]{%
  \expandafter\let\csname old#1\expandafter\endcsname\csname #1\endcsname
  \expandafter\let\csname oldend#1\expandafter\endcsname\csname end#1\endcsname
  \renewenvironment{#1}%
     {\linenomath\csname old#1\endcsname}%
     {\csname oldend#1\endcsname\endlinenomath}}%
\newcommand*\patchBothAmsMathEnvironmentsForLineno[1]{%
  \patchAmsMathEnvironmentForLineno{#1}%
  \patchAmsMathEnvironmentForLineno{#1*}}%
\AtBeginDocument{%
\patchBothAmsMathEnvironmentsForLineno{equation}%
\patchBothAmsMathEnvironmentsForLineno{align}%
\patchBothAmsMathEnvironmentsForLineno{flalign}%
\patchBothAmsMathEnvironmentsForLineno{alignat}%
\patchBothAmsMathEnvironmentsForLineno{gather}%
\patchBothAmsMathEnvironmentsForLineno{multline}%
}

\usepackage{hyperref}%
\hypersetup{%
  breaklinks=true,%
  linktocpage=true,
  colorlinks=true,%
  linkcolor=black,%
  citecolor=blue,%
  urlcolor=black,%
  pdftitle={},%
  pdfsubject={},%
  pdfauthor={Axel Finke},%
  pdfkeywords={},%
  pdfproducer={}%
  }%

\usepackage{ifthen}%
\usepackage{mfirstuc}
\usepackage{xkeyval}%
\usepackage{xfor}%
\usepackage{amsgen}%
\usepackage{etoolbox}%
\usepackage{datatool-base}%

\usepackage[%
  xindy,%
  style=long,%
  toc=true,%
  nonumberlist,%
  acronym,%
  acronymlists={abbreviations},%
  hyperfirst=true
  ]{glossaries}%

\newacronym{PMCMC}{PMCMC}{particle Markov chain Monte Carlo}%
\newacronym{PF}{PF}{particle filter}%
\newacronym{EHMM}{EHMM}{embedded hidden Markov models}%
\newacronym{HMM}{HMM}{hidden Markov model}%
\newacronym{CPF}{CPF}{conditional particle filter}%
\newacronym{PG}{PG}{particle Gibbs}%
\newacronym{PGBS}{PG-BS}{particle Gibbs sampler with backward sampling}%
\newacronym{PGAS}{PG-AS}{particle Gibbs sampler with backward sampling}%
\newacronym{SQMC}{SQMC}{sequential quasi Monte Carlo}%
\newacronym{RQMC}{RQMC}{randomised quasi Monte Carlo}%
\newacronym[user1={ancestor-sampling}]{AS}{AS}{ancestor sampling}%
\newacronym[user1={backward-sampling}]{BS}{BS}{backward sampling}%
\newacronym{PDF}{PDF}{probability density function}%
\newacronym{IID}{IID}{independent and identically distributed}%
\newacronym{MCMC}{MCMC}{Markov chain Monte Carlo}%
\newacronym{MH}{MH}{Metropolis--Hastings}%
\newacronym{ESS}{ESS}{effective sample size}%
\newacronym{CDF}{CDF}{cumulative distribution function}%
\newacronym{SMC}{SMC}{sequential Monte Carlo}%
\newacronym{CSMC}{CSMC}{conditional sequential Monte Carlo}%
\newacronym{EPSRC}{EPSRC}{Engineering and Physical Sciences Research Council}%
\newacronym{LW}{LW}{Liu~\&~West}%
\newacronym{CLT}{CLT}{central limit theorem}%
\newacronym{WLLN}{WLLN}{weak law of large numbers}%
\newacronym{IACT}{IACT}{integrated autocorrelation time}%
\newacronym{EM}{EM}{expectation--maximisation}%
\newacronym{ML}{ML}{maximum-likelihood}%
\newacronym{GWAS}{GWAS}{genome-wide association studies}%
\newacronym{WGBS}{WGBS}{whole-genome bisulphite sequencing}%
\newacronym{RRBS}{RRBS}{reduced representation bisulphite sequencing}%
\newacronym{SVA}{SVA}{surrogate-variable analysis}%
\newacronym{DMR}{DMR}{differentially methylated region}%
\newacronym{DMP}{DMP}{differentially methylated position}%
\newacronym{DMC}{DMC}{differentially methylated COMET}%
\newacronym[firstplural={blocks of comethylation (COMETs)}]{COMET}{COMET}{block of comethylation}%
\newacronym{UMR}{UMR}{unmethylated region}%
\newacronym{LMR}{LMR}{lowly methylated region}%
\newacronym{PMD}{PMD}{partially methylated domains}%
\newacronym{DNA}{DNA}{deoxyribonucleic acid}%
\newacronym{CpG}{CpG}{cytosine-guanine dinucleotide}
\newacronym{EWAS}{EWAS}{epigenome-wide association studies}%
\newacronym{FDR}{FDR}{false discovery rate}%
\newacronym{FNR}{FNR}{false non-discovery rate}%
\newacronym{FDP}{FDP}{false discovery proportion}%
\newacronym{FNP}{FNP}{false non-discovery proportion}%
\newacronym{mFDR}{mFDR}{marginal false discovery rate}%
\newacronym{mFNR}{mFNR}{marginal false non-discovery rate}%
\newacronym{TP}{TP}{true positives}%
\newacronym{DSS}{DSS}{dispersion shrinkage for sequencing}
\newacronym{DMRSEQ}{DMRSEQ}{-----} 

\usepackage{csquotes}%
\usepackage{natbib}
\setcitestyle{authoryear}
\qedsymbol{\ensuremath{\Box}}
\theoremstyle{plain}%

\theoremheaderfont{\normalfont\bfseries\rmfamily} 
\theorembodyfont{\normalfont\slshape}%
\theoremseparator{.}

\theorembodyfont{\normalfont}%
\newmdtheoremenv[
 ntheorem=true,
 skipbelow = .6\baselineskip plus 1ex minus 1ex,
 skipabove = .6\baselineskip plus 1ex minus 1ex,
 innerleftmargin = 0pt,
 innerrightmargin = 0pt,
 leftline = false,
 rightline = false,
 needspace = 5ex 
]{framedAlgorithm}[theorem]{Algorithm}

\theoremstyle{empty}%
\theoremheaderfont{} 
\theoremstyle{nonumberplain}%
\theorembodyfont{\upshape \rmfamily}%
\theoremheaderfont{\upshape \rmfamily\bfseries}%
\theoremsymbol{\ensuremath{_\Box}}
\usepackage[textwidth=2.5cm, textsize=scriptsize]{todonotes}

\usepackage{tabularx, booktabs}
\usepackage{multirow}

\deffootnote[1em]{1em}{1em}{\textsuperscript{\thefootnotemark}}

\begin{document}

\nolinenumbers

\title{\Large{A Bayesian framework for genome-wide inference of DNA methylation levels}}

\renewcommand{\thefootnote}{\fnsymbol{footnote}}
\author{\large{
Marcel Hirt$^{*,\dag}$, 
Axel Finke$^\ddag$,
Alexandros Beskos$^{\dag,\mathsection}$,
Petros Dellaportas$^{\dag, \mathparagraph}$,
Stephan Beck$^{\|}$,
Ismail Moghul$^{\|}$,
Simone Ecker$^{\|}$
}}

\date{\large{\today}}

\maketitle

\glsunset{DNA}
\glsunset{DMRSEQ}


\begin{abstract}
    \noindent{}\gls{DNA} methylation is an important epigenetic mark that has been studied extensively for its regulatory role in biological processes and diseases. \Glsdesc{WGBS} allows for genome-wide measurements of DNA methylation up to single-base resolutions, yet poses challenges in identifying significantly different methylation patterns across distinct biological conditions. We propose a novel methylome change-point model which describes the joint dynamics of methylation regimes of a case and a control group and benefits from taking into account the information of neighbouring methylation sites among all available samples. We also devise particle filtering and smoothing algorithms to perform efficient inference of the latent methylation patterns. We illustrate that our approach can detect and test for very flexible differential methylation signatures with high power while controlling Type-I error measures.
\end{abstract}

\makeatletter
\addtocounter{footnote}{1} \footnotetext{
School of Physical and Mathematical Sciences, NTU, Singapore}
\addtocounter{footnote}{1} \footnotetext{ 
Department of Statistical Science, UCL, UK}
\addtocounter{footnote}{1} \footnotetext{
Department of Mathematical Sciences, Loughborough University, UK}
\addtocounter{footnote}{1} \footnotetext{
Alan Turing Institute for Data Sciences, UK}
\addtocounter{footnote}{1} \footnotetext{
Department of Statistics, Athens University of Economics and Business,  Greece}
\addtocounter{footnote}{1} \footnotetext{
Cancer Institute, UCL, UK}
\makeatother

\setcounter{footnote}{0}
\renewcommand{\thefootnote}{\arabic{footnote}}

\glsreset{WGBS}

\section{Introduction}

\subsection{DNA methylation}

Methylation in mammals is an epigenetic modification of \gls{DNA} that adds a methyl group at position C5 in the context of \emph{\glspl{CpG}} \citep{bird2002dna}. Methylation is known to be associated with organismal development, aging and progression of human diseases such as cancer \citep{robertson2005dna}. High-throughput sequencing techniques provide high-resolution methylation profiles on a genome-wide scale, with \emph{\gls{WGBS}} becoming a gold-standard technique for methylation studies \citep{bock2012analysing}, due to its single-base coverage and high accuracy. The diploid human epigenome has more than $10^7$ \gls{CpG} sites within its (more than $10^8$) cytosines, making a statistical analysis of such data sets extremely challenging undertaking, particularly in the context of \emph{\gls{EWAS}} which aim to link epigenetic variations to particular phenotypes for example in cases/control studies \citep{rakyan2011epigenome}. For illustration, \gls{DNA} \emph{hypomethylation}, i.e.\ relative undermethylation within promoter regions, has been shown to inactivate certain tumor-suppressor genes, while \gls{DNA} \emph{hypermethylation}, i.e.\ global overmethylation, can induce genomic instability, thus contributing to cell transformation \citep{kulis2010dna}.

\subsection{Existing approaches}

A large number of approaches have been suggested to detect \emph{\glspl{DMP}} or \emph{\glspl{DMR}} between case and control groups, i.e.\ positions (single \gls{CpG} sites) or regions (consisting of multiple neighbouring \gls{CpG} sites) in which the methylation level differs significantly between the case and control groups -- see the review in \citet{shafi2018survey}.

A commonly used approach suggested in \citet{hansen2012bsmooth} smoothes the methylation values and then tests for group differences using $t$-tests for each site, but does not allow for a control of Type-I error rates when performing multiple-hypothesis testing across sites or regions, while also not being able to handle missing reads.

Different beta-binomial models have been suggested \citep{burger2013identification, feng2014bayesian, park2014methylsig,sun2014moabs, wu2015detection}, but they tend to allow for only limited spatial dependence, as do most approaches relying on logistic regressions \citep{akalin2012methylkit}, linear mixed models \citep{jaffe2012bump} or established statistical tests \citep{stockwell2014dmap}.

Generalised linear models \citep{korthauer2019detection, kapourani2021scmet, halla2019luxus}, quasi-binomical mixed models \citep{zhao2021detecting} or wavelet-based functional mixed models \citep{denault2021detecting} have been suggested to test for \glspl{DMR}, but these often require that such regions of interest have already been discovered, or they are not scalable to genome-wide inference \citep{lee2016identification}. 

Further models of \gls{DNA} methylation have been suggested based on a one-dimensional Ising model \citep{jenkinson2017potential,jenkinson2018information}, a latent Gaussian-field model \citep{rackham2017bayesian, hubin2020bayesian} or a spline-based functional-regression model \citep{shokoohi2021identifying}. However, due to high computational complexity, estimating such models requires 
\begin{enumerate*}[label = (\alph*)]
    \item partitioning of the genome into small sub-regions, or
    \item approximate inference techniques. 
\end{enumerate*}



\subsection{Contributions}
\label{subsec:contributions}


We propose a novel statistical methodology for discovering methylation patterns based on a class of Bayesian change-point models. Specifically, our contributions are threefold.
    \paragraph{Single-group inference.} In Section~\ref{sec:single_group}, we present a first model for discovering and retrieving methylation patterns when only epigenetic samples for a single group are available. This methodology provides the following advantages.
    \begin{enumerate}
      \item \label{enum:advantages:1}  \textit{Discovery of general methylation patterns.} Our method can be used to probabilistically segment the methylome into regions of interest, whilst also incorporating domain/expert knowledge as to what different methylation regimes are known or expected to look like. In particular, the methylation patterns can be specified very generally. In contrast, most existing methods can only detect changes in the mean.
      
      \item \textit{Accounting for spatial correlation.} Our model accounts for the fact that methylation is generally spatially correlated but can also change abruptly. In contrast, widely-used existing methods such as BSmooth \citep{hansen2012bsmooth} tend to smooth out any abrupt changes.
      
      
      \item \textit{Principled imputation of missing data.} Our methodology also works in the presence of missing reads. That is, our method probabilistically imputes missing methylation measurements based on reads from nearby \gls{CpG} sites.
      
      \item \label{enum:advantages:4} \textit{Works with a single sample.} Our methodology still works even if we only have a single replicate. In contrast, some existing methods require at least two samples.
      
      \item \label{enum:advantages:5} \textit{Efficient genome-wide inference.} Our methodology allows efficient inference on a genome-wide scale by relying on \emph{\gls{SMC}} algorithms \citep{fearnhead2007online,  fearnhead2009bayesian, caron2012online, yildirim2013online} whose computational complexity grows only linearly with the number of \gls{CpG} sites -- an attribute regarded as a great challenge for previous approaches.
    \end{enumerate}
    
    \paragraph{Case--control inference.} In Section~\ref{sec:case_control} we extend the model to enable the discovery of differences in methylation patterns between epigenetic samples in a \emph{case} group and a \emph{control} group. Our proposed methodology directly inherits all the Advantages~\ref{enum:advantages:1}--\ref{enum:advantages:4} from the single-group scenario. In addition, it provides the following advantages.
    \begin{enumerate}
      \item \textit{Joint modelling of cases and controls.} Our approach obtains greater statistical efficiency by modelling the case and control groups \emph{simultaneously}. In contrast, some existing methods lose efficiency by fitting independent models to each group.
      
      \item \textit{Works with a single case and single control sample.} Our methodology can still be used even if only a single sample from each group is available; but it also scales to large cohorts and population-wide studies.
      
      \item \textit{Direct Bayesian inference of regions of interest.} Our model includes a parameter which specifies whether the case and control group have different methylation patterns at a given \gls{CpG} site. We can thus directly perform Bayesian inference on regions of interest by approximating the posterior distribution of these parameters.
      
       \item \textit{Efficient genome-wide inference.} Although part of the \gls{SMC} methodology from the single-group scenario is not applicable to the case--control model, we devise a different \gls{SMC} algorithm based around the \emph{discrete particle filter} from \citet{fearnhead1998sequential, fearnhead2003online} which again has a computational complexity that is linear in the number of \gls{CpG} sites and thus permits genome-wide inference.
    \end{enumerate}

    \paragraph{Discovery of methylation signatures.} 
    In Section~\ref{sec:identification}, we illustrate how to leverage the data-driven inference results from the single-group or case--control setting for a hypothesis-based discovery of flexible methylation signatures. This principled decision framework provides the following advantages.
    \begin{enumerate}
        \item \textit{Improved power for a given \gls{FDR} target.}
        A series of works in multiple testing  \citep{sun2007oracle, sun2009large, sun2011multiple, sun2015false} have established \emph{optimal} testing procedures that maximize the power subject to a constraint on the \gls{FDR} in case--control scenarios. These procedures are based on the posterior probability of the latent signal, and thereby automatically take into account the spatial dependency of the underlying signal process.
        \item \textit{Effective \gls{FDR} control.}
        Our inference procedures based on \gls{SMC} provide an effective approximation to the optimal procedure that tightly controls the \gls{FDR}. In contrast, multiple-testing procedures based on $p$-values can be overly conservative or challenging to derive under dependence assumptions.
        \item \textit{Capacity to test for flexible methylation signal at different spatial resolutions.} Our framework can be used to test for various methylation-pattern hypotheses, not just concerning methylation means or variances, and at different resolutions: From single sites to regions that can be pre-defined based on genomic annotations or constructed using our change-point model. Conventionally, this required the application of multiple, unlinked analysis methods.
    \end{enumerate}




\subsection{Related literature}

Our approach is related to existing approaches based around \emph{\glspl{HMM}} such as \citet{molaro2011sperm, kuan2012integrating, saito2014bisulfighter, yu2016hmm, sun2016hmm, shen2017detect, shokoohi2019hidden}.  However, we stress that even the relatively simple single-group model proposed in Section~\ref{sec:single_group} is more general than a \gls{HMM}. Indeed, the single-group model change-point model is closely related to a hidden-semi Markov model. Such models were applied to infer differences in methylation patterns in \citet{du2014biomvrhsmm}. Importantly, none of these existing approaches simultaneously share all the advantages listed in Section~\ref{subsec:contributions}.

Change-point detection methods for count data have been developed and applied for peak detection and segmentation on ChIP-seq data sets, see for example \citep{hocking2015peakseg, hocking2020constrained, hocking2022generalized} using dynamic programming ideas, whilst hidden Markov models have been suggested to annotate for example chromatin states \citep{ernst2017chromatin}. However, such approaches are not immediately applicable for annotating or detecting differential methylation patterns for \gls{WGBS} data.

\section{Single-group methylome model}
\label{sec:single_group}

In this section, we present a relatively simple change-point model which can be used to discover methylation patterns when only epigenetic samples for a single group are available. The model is useful in its own right because it all allows us to model and discover more general methylation patterns than existing approaches which can only distinguish methylation patterns that affect the mean or variance. Importantly, our approach permits a very efficient inference of such methylation patterns at a computational cost that only grows linearly in the number of \gls{CpG} sites. The model also forms the main building block for our model for case--control scenarios which will be presented in the next section.

\subsection{Data}

Consider $S \geq 1$ available epigenetic samples (all belonging to the same group) and let $t = 1,\dotsc, T$ be the numbering of the \gls{CpG} sites in some chromosome. Typically, $T$ is of order $10^6$. Observations are denoted $(y_t)_{t \geq 1}$, where $y_t \coloneqq y_{t,1:S}$. Here, $y_{t,s} \in \naturals_0 \coloneqq \naturals \cup \{0\}$ denotes the number of methylated reads at the $t$th \gls{CpG} site associated with the $s$th sample. We assume that the corresponding total number of reads, $n_{t,s} \in \naturals_0$, is known; we write $n_t \coloneqq n_{t,1:S}$. Throughout, we will sometimes also refer to the index $t$ as representing ``time'' because our inference algorithms will browse through the \gls{CpG} sites in a chromosome from left to right.

\subsection{Likelihood}
To take into account the variability in the reads, we assume that
\begin{align}
    p(y_{1:T}|\pi_{1:T}, x_{1:T}, \theta)
    = \prod_{t=1}^T \prod_{s=1}^S p(y_{t,s}|\pi_{t,s}, \theta), 
\end{align}
where $\theta$ is a collection of unknown model parameters (to be specified later) and where
\begin{align}
   p(y_{t,s}|\pi_{t,s}, \theta) \coloneqq \dBin(y_{t,s}; n_{t,s}, \pi_{t,s}).
\end{align}
Here, $\dBin(x, n, \pi)$ denotes the probability mass function of a Binomial distribution with size parameter $n$ and probability $\pi$ evaluated at $x$. Furthermore, $\pi_{t,s}$ is a probability which can be interpreted as the unobserved proportion of methylated alleles over all cells in the $s$th sample at the $t$th \gls{CpG} site. As usual, we write $\pi_t \coloneqq \pi_{t,1:S}$. A~priori, we model these probabilities as
\begin{align}
    p(\pi_{1:T}| x_{1:T}, \theta)
    = \prod_{t=1}^T \prod_{s=1}^S p(\pi_{t,s} | x_t, \theta),
\end{align}
where
\begin{align}
  p(\pi_{t,s} | x_t, \theta)
  \coloneqq \dBeta(\pi_{t,s}; \alpha_{r_t}, \beta_{r_t}). \label{eq:distribution_of_pi}
\end{align}
Here, $\dBeta(x;\alpha,\beta)$ denotes density of a beta distribution with parameters $\alpha, \beta > 0$, evaluated at $x$. The random variable $r_t \in [R]$ is a \emph{regime indicator}. We assume that there are a finite number, $R$, of possible \emph{regimes}, i.e.\ $r_t \in [R]$, where we use the notation $[n] \coloneqq \{1,2,\dotsc,n\}$. In \eqref{eq:distribution_of_pi}, each regime is associated with a different pair of parameters $(\alpha_r, \beta_r)$ and thus implies a different prior distribution on the unobserved probabilities $\pi_{t,s}$. In particular, in the single-group setting assumed here, the Regime~$r_t$ at the $t$th site is the same over all samples $s \in [S]$. Here and below, $r_t$ is a component of the latent vector $x_t$ (other components of $x_t$ are specified later).

\subsection{Integrating out the latent probabilities} 

Conditional on the latent vectors $x_{1:T}$ -- recall that these include the unobserved regime indicators $r_{1:T}$ -- we can analytically integrate out all the probabilities $\pi_{1:T}$ to obtain the product of beta-binomial likelihoods
\begin{align}
  p(y_{1:T}|x_{1:T}, \theta)
   & = \int p(y_{1:T}|\pi_{1:T}, x_{1:T}, \theta) p(\pi_{1:T}| x_{1:T}, \theta) \intDiff \pi_{1:T} 
   = \prod_{t=1}^{T} g_{\theta,t}(x_t), \label{eq:potential_function_integral}
\end{align}
where, on the right-hand-side, we have dropped the observations from the expression for the  quantities $g_{\theta,t}(x_t)$. The latter are determined analytically as
\begin{align}
  g_{\theta,t}(x_t) 
  & \coloneqq \prod_{s = 1}^{S} \int_0^1 \dBin(y_{t,s}; n_{t,s}, \pi) \dBeta(\pi; \alpha_{r_t}, \beta_{r_t}) \intDiff \pi\\
  & = \prod_{s = 1}^{S} \dBetaBin(y_{t,s}; n_{t,s}, \alpha_{r_t}, \beta_{r_t}).
\end{align}
Here, $\dBetaBin(y; n, \alpha, \beta)$ denotes the probability mass function of a beta-binomial distribution, i.e.\
\begin{align}
  \MoveEqLeft \dBetaBin(y; n, \alpha, \beta)\\ 
  & \coloneqq \binom{n}{y} \frac{\mathrm{B}(y + \alpha, n-y+\beta)}{\mathrm{B}(\alpha,\beta)}\\
  & = \frac{\Gamma(n+1)}{\Gamma(y+1)\Gamma(n-y+1)} \frac{\Gamma(y+\alpha)\Gamma(n-y+\beta)}{\Gamma(n+\alpha+\beta)} \frac{\Gamma(\alpha + \beta)}{\Gamma(\alpha)\Gamma(\beta)},
\end{align}
where $\mathrm{B}(\ccdot, \ccdot)$ and $\Gamma(\ccdot)$ denote the beta and gamma functions, respectively.

The likelihood $p(y_{1:T}|x_{1:T}, \theta)$ is properly defined in the case $n_{t,s} = 0$ for all $s \in [S]$, that is when no sample has any reads at the $t$th \gls{CpG} site. The change-point model thereby naturally allows the imputation of the latent methylation state. However, we leave an empirical comparison against alternative imputation methods \citep[such as][]{ernst2015large, angermueller2017deepcpg} for future work. 

\subsection{Methylation regimes}

Based on empirical evidence and expert-knowledge elicitation\footnote{The number of regimes and the regime-specific parameters can be changed and adjusted to accommodate particular biological research questions, different technical aspects and sequencing techniques. By varying the mean and standard deviation parameter, different methylation patterns can be described using a beta-binomial density -- for instance also bi-stable regimes with bi-modal densities as considered in \citet{jenkinson2017potential, jenkinson2018information} -- while being relatively parsimonious. For visualisation of methylation regimes with slightly perturbed mean or variance parameters, see Figure~\ref{fig:regimes_simulation}.}, we envisage six different model regimes for the complete genome. These are summarised in Table~\ref{tab:regimes}. To facilitate interpretation, we present these regime-specific parameters not in the standard parametrisation $(\alpha_r, \beta_r)$ for a beta law but in the form of its mean $\mu_r$ and standard deviation $\sigma_r$. The standard parameters can then be recovered via the relationship
\begin{align}
 \alpha_r = \mu_r \nu_r, \quad 
 \beta_r = (1-\mu_r)\nu_r, \quad \text{where} \quad 
 \nu_r \coloneqq \frac{\mu_r(1-\mu_r)}{\sigma_r^2} - 1,
\end{align}
for $\sigma_r < \sqrt{\mu_r(1-\mu_r)}$.

\begin{table}[htb]
 \centering
 \caption{The $R = 6$ regimes in the single-group scenario.}
 \begin{tabular}{@{}llSS@{}}
  \toprule
  {Regime $r$} & {Description} & {Mean $\mu_r$} & {Std dev.~$\sigma_r$} \\\midrule
  1 & very large mean, small std~dev.\     & 0.95 & 0.05\\
  2 & very small mean, small std~dev.\     & 0.05 & 0.05\\
  3 & large mean, moderate std~dev.\       & 0.8 & 0.1\\
  4 & small mean, moderate std~dev.\       & 0.2 & 0.1\\
  5 & mean around 0.5, moderate std~dev.\  & 0.5 & 0.1\\
  6 & `chaotic' (i.e.\ uniform on $[0,1]$) & 0.5 & {$1/\sqrt{12}$}\\
\bottomrule
 \end{tabular}
  \label{tab:regimes}
\end{table}

\begin{figure}[htb!]
 \centering
 \includegraphics[scale=0.65, trim=0.5cm 0.5cm 0cm 0.5cm]{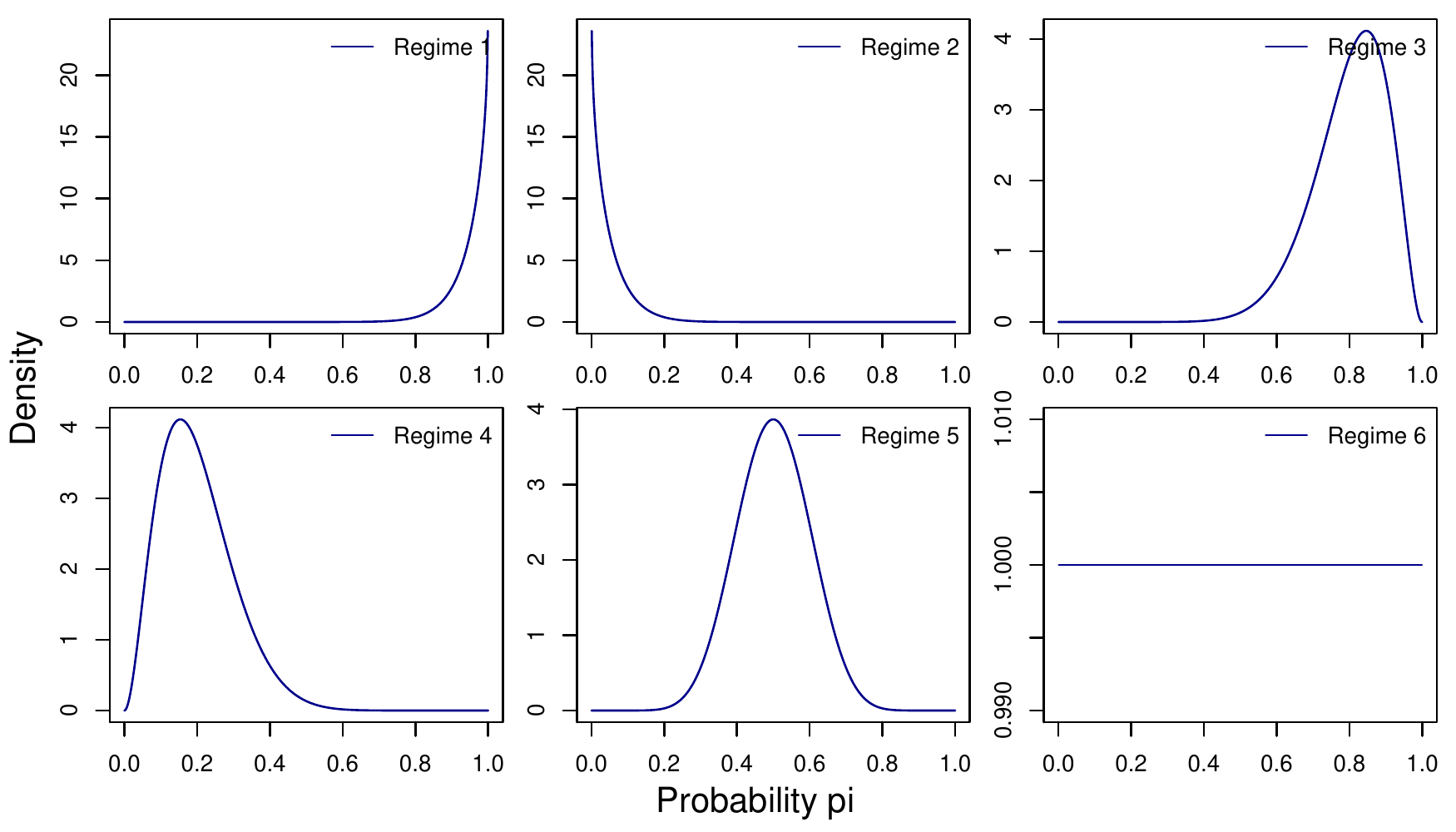}
 \caption{Densities of the beta distribution for Regimes~1--6, under the hyperparameter choices shown in Table~\ref{tab:regimes}.
 }
 \label{fig:beta_config8}
\end{figure}

\subsection{Change points}
\label{subsec:model_single_case}

We now specify the change-point model, i.e.\ the model which governs how the regime indicators evolve as we browse through the \gls{CpG} sites from left to right.

To that end, we follow the setup of \citet{fearnhead2007online, caron2012online, yildirim2013online} and include a \emph{sojourn time} $d_t$ into $x_t$ which specifies for how long (i.e.\ for how many sites leading up to site~$t$ from the left) the model has already been in Regime~$r_t \in [R]$. More formally, we write
\begin{align}
  x_t = (d_t, r_t).
\end{align}
Then, at the $t$th \gls{CpG} site in the chromosome, $d_t$ denotes the distance to the most recent change point. We assume that a~priori, $(x_t)_{t \in \naturals}$ is a a latent Markov chain taking values in
\begin{align}
  \calX \coloneqq \naturals \times [R]. 
\end{align}
This Markov structure will later facilitate the use of computationally effective particle filtering algorithms for the model fitting; for a similar modelling approach allowing message passing computations at the point of calibration, see \citet{adams2007bayesian}. 

With some abuse of notation, let $\delta_x$ be the point mass centred at $x$, i.e.\ $\delta_x(y) = \ind\{x = y\}$ where $\ind$ is the indicator function. A~priori, the Markov chain $(x_t)_{t \in \naturals}$ has initial distribution $f_{\theta,1}(x_1) \coloneqq \delta_1(d_1) \nu_\theta(r_1)$ -- for some distribution $\nu_\theta$ on $[R]$ -- and transition kernels
\begin{align}
 f_{\theta,t+1}(x_{t+1}|x_{t})
 & \coloneqq \rho_{\theta,t+1}(x_{t}) \delta_1(d_{t+1}) P_\theta(r_{t+1}|r_{t})\\
 & \quad\quad + (1 - \rho_{\theta, t+1}(x_{t})) \delta_{d_{t}+1}(d_{t+1})\delta_{r_t}(r_{t+1}).
\end{align}
Here, given a current state $x_{t}=(d_{t},r_{t})$: 

\begin{enumerate}
  \item $\rho_{\theta,t+1}(x_{t}) \in [0,1]$ is the probability of occurrence of  a change point, after having spent $d_{t}$ time steps in regime $r_{t}$. Following \citet{caron2012online}, we select a probability mass function $h_{\theta, r_t}$ with support on $\naturals$ (or on some subset thereof) such that $h_{\theta, r_t}(d_t)$ constitutes the prior probability over the sojourn time, $d_t$, in the $r_t$th regime. Thus, 
  \begin{align}
   \rho_{\theta,t+1}(x_{t})
   & \coloneqq 
   \frac{h_{\theta, r_t}(d_{t})}{1 - H_{\theta,r_t}(d_{t}-1)},
 \end{align}
  where $H_{\theta,r_t}(l) \coloneqq \sum_{k=1}^l h_{\theta, r_t}(k)$ denotes the cumulative distribution function associated with $h_{\theta, r_t}$.

  \item $P_\theta(r_{t+1}|r_{t})$ is the $r_{t+1}$th entry in the $r_{t}$th row of an $R \times R$ matrix of transition probabilities determining the new generative model at the occurrence of a change point. To rule out trivial change points, we assume that this matrix has zeros on its diagonal.
\end{enumerate}
Conditional on the model parameters $\theta$, the posterior distribution of interest is then given by
\begin{align}
 p(x_{1:T}|y_{1:T}, \theta) \propto \prod_{t=1}^T \gamma_{\theta, t}(x_{t-1}, x_t),
\end{align}
where $\gamma_{\theta,t}(x_{t-1}, x_t) \coloneqq f_{\theta,t}(x_t|x_{t-1}) g_{\theta,t}(x_t)$.

\subsection{Model parameters} 
\label{subsec:model_params_single_case}

For any regime $r \in [R]$, we specify the prior distribution over the number and location of the change points as follows. Following \citet{caron2012online},  we take $h_{\theta, r}(d) \coloneqq \dNegBin(d-u_r;\kappa_r,\omega_r) \ind\{d \geq u_r\}$ to be the probability mass function of a shifted negative-binomial distribution, where $u_r \geq 1$ is the shift, and
\begin{align}
 \dNegBin(z;\kappa,\omega) \coloneqq 
 \frac{\Gamma(z+\kappa)}{\Gamma(\kappa)\Gamma(z+1)}
 \omega^{z}  (1-\omega)^{\kappa},
\end{align}
for $z \in \naturals_0$. For any $r \in [R]$, we assume that the shifts $u_r \geq 1$ are known. We further follow \citet{caron2012online} and set $\kappa_1 = \dotsc = \kappa_R = 2$ in our experiments although we note that these parameters could also be estimated from the data.

The parameters in the model which remain unknown (and which we thus need to estimate from the data) are then:
\begin{enumerate}
    \item the entries of the transition matrix $P_\theta(r'|r)$;
    \item the regime-specific `success probability' parameters $\omega_1, \dotsc, \omega_R$.
\end{enumerate}
In total, we thus have $R^2$ unknown parameters collected in the vector $\theta \coloneqq \theta_{1:R^2} \in \reals^2$ after applying some bijective transformations outlined in Appendix~\ref{app:subsec:single-group:parametrisation}.

\subsection{Inference}
\label{subsec:single_group_inference}
We now discuss how one can efficiently perform inference in the above-mentioned single-group change-point model using \gls{SMC} methods, a.k.a.\ particle filters. Specifically, our approach is based around the so-called particle filter for change-point models introduced in \citet{fearnhead2007online}. Briefly, our methodology comprises three elements (see Appendix~\ref{app:sec:single_group} for details):
\begin{enumerate}
  \item \textbf{Filtering.} We use the particle filter for change-point models from \citet{fearnhead2007online} \citep[see also][]{whiteley2011bayesian, yildirim2013online} to approximate the `filtering' distributions $p(x_t|y_{1:t}, \theta)$ for all sites $t \in [T]$. More details are given in Appendix~\ref{app:subsec:single_group:filtering}.
  
  \item \textbf{Smoothing.} Our main interest is in the posterior distributions of the latent variables, $p(x_t|y_{1:T}, \theta)$ and, in particular, of the regime indicators, $p(r_t|y_{1:T}, \theta)$, for all sites $t \in [T]$. We use the adaptive-lag smoother from \citet{alenlov2019particle} to efficiently approximate these quantities. More details are given in Appendix~\ref{app:subsec:single_group:smoothing}.
  
  \item \textbf{Parameter estimation.} The model parameters $\theta$ are unknown. We use an online gradient-ascent algorithm \citep{poyiadjis2005maximum, caron2012online} to approximate their maximum-likelihood estimate (separately for each chromosome). More details are given in Appendix~\ref{app:subsec:single_group:parameter_estimation}.
\end{enumerate}
We stress that all these methods only need to browse through the data once. Thus, their computational complexity is linear in the total number of \gls{CpG} sites, $T$. Since $T$ is of order $10^6$ for each chromosome, this constitutes a crucial advantage of our \gls{SMC} methodology over other methods, such as \gls{MCMC} algorithms.

\section{Case--control methylome model}
\label{sec:case_control}

In this section, we describe an extended change-point model for modelling and discovering differences in methylation patterns between epigenetic samples from a \emph{case} group and a \emph{control} group. The observations in each group will be modelled as a change-point model, i.e.\ we now have two change-point models: one for the control group and one for the case group. However, these two models \emph{are not independent} -- the methylation states for the case group generally depend on the methylation signal of the control group. More specifically, for most sites, there is no difference between the case group and the control group. However, occasionally, the case group `splits off' and evolves separately; the model includes latent binary indicators $z_t$ which indicate whether or not the two groups are merged or split at Site~$t$. The posterior distribution of these indicators then lets us assess, in a principled manner with full uncertainty quantification, whether the case and control group have different methylation patterns at a given collection of \gls{CpG} sites; and, importantly, we present a \gls{SMC} algorithm which allows us to efficiently approximate the posterior distribution of these indicators at a cost that again only grows linearly in the total number of \gls{CpG} sites.

\subsection{Data and latent variables}

We now have observations for two groups: \emph{controls} and \emph{cases}. We use the convention that the `bar'-accented quantities are associated with the `control' group whereas `tilde'-accented quantities are associated with the `case' group. That is, for the respective groups, and for the $t$th \gls{CpG} site,
\begin{enumerate}
 \item $\yCon_t \coloneqq \yCon_{t,1:\SCon}$ and $\yCas_t \coloneqq \yCas_{t,1:\SCas}$ are the observed numbers of methylated reads over all samples,
 \item $\nCon_t \coloneqq \nCon_{t,1:\SCon}$ and $\nCas_t \coloneqq \nCas_{t,1:\SCas}$ are the total number of reads over all samples,
 \item $\piCon_t\coloneqq \piCon_{t,1:\SCon}$ and $\piCas_t \coloneqq \piCas_{t,1:\SCas}$ are the `methylation probabilities', i.e.\ these can again be interpreted as the unobserved proportions of methylated alleles over all cells over all samples,
 \item $\xCon_{t} \coloneqq (\dCon_{t},\rCon_{t})$ and $\xCas_{t} \coloneqq (\dCas_{t},\rCas_{t})$ are the unobserved sojourn times and regime indicators.
\end{enumerate}
To keep the notation concise, we also write 
\begin{align}
 \yAll_t  & \coloneqq (\yCon_t, \yCas_t),\\
 \nAll_t  & \coloneqq (\nCon_t, \nCas_t),\\
 \piAll_t & \coloneqq (\piCon_t, \piCas_t),\\
 \xAll_t  & \coloneqq (z_t, \xCon_t, \xCas_t) = (z_t, \dCon_{t},\rCon_{t},\dCas_{t},\rCas_{t}),
\end{align}
where $z_t \in \{0, 1\}$ denotes an additional binary latent variable which governs the dependence between the two groups. Its role will be made precise below. Loosely speaking, if $z_t = 1$, then we assume that the methylation patterns in both groups are identical, i.e.\ the case group evolves exactly as the control group. If $z_t = 0$, then the methylation pattern in the case group is able to diverge from the control group.

\subsection{Likelihood}

As before, we use a beta-binomial model for the number of methylated reads in each group and sample, i.e.\ we assume that
\begin{align}
  \MoveEqLeft p(\yAll_{1:T}|\piAll_{1:T}, \xAll_{1:T}, \theta)\\
  &= \prod_{t=1}^T \biggl[\prod_{s=1}^{\SCon} \dBin(\yCon_{t,s}; \nCon_{t,s}, \piCon_{t,s})\biggr]\biggl[\prod_{s=1}^{\SCas}\dBin(\yCas_{t,s}; \nCas_{t,s}, \piCas_{t,s})\biggr].
\end{align}
The main difference with the single-group scenario is that probabilities $\piCon_{t,s}$ and $\piCas_{t,s}$ are now not necessarily \emph{identically} distributed conditionally on the latent regimes. That is, we model
\begin{align}
  p(\piAll_{1:T}|\xAll_{1:T},\theta)
  \coloneqq \prod_{t=1}^T \biggl[\prod_{s=1}^{\SCon} \dBeta(\piCon_{t,s}; \alpha_{\rCon_t},\beta_{\rCon_t}) \biggr] \biggl[\prod_{s=1}^{\SCas} \dBeta(\piCas_{t,s}; \alpha_{\rCas_t},\beta_{\rCas_t}) \biggr], \label{eq:case_control_pi:default}
\end{align}
where we stress that the regime indicators at Site~$t$ can be different. 

\subsection{Integrating out the latent probabilities} 

As in the single-group scenario, we can analytically integrate out all the probabilities $\pi_{1:T}$ to obtain the product of beta-binomial likelihoods
\begin{align}
  p(\yAll_{1:T}|x_{1:T}, \theta)
   & = \int p(\yAll_{1:T}|\piAll_{1:T}, \xAll_{1:T}, \theta) p(\piAll_{1:T}|\xAll_{1:T},\theta) \intDiff \piAll_{1:T}  = \prod_{t=1}^{T} \gAll_{\theta,t}(\xAll_t),
\end{align}
where, writing 
\begin{align}
 \gCon_{\theta,t}(\xCon_t) 
 &\coloneqq \prod_{s = 1}^{\SCon} \dBetaBin(\yCon_{t,s}; \nCon_{t,s}, \alpha_{\rCon_t}, \beta_{\rCon_t}),\\[0.6ex]
 \gCas_{\theta,t}(\xCas_t)
 & \coloneqq \prod_{s = 1}^{\SCas} \dBetaBin(\yCas_{t,s}; \nCas_{t,s}, \alpha_{\rCas_t}, \beta_{\rCas_t}),
\end{align}
we have set
\begin{align}
 \gAll_{\theta,t}(\xAll_t)
 & \coloneqq 
 \gCon_{\theta,t}(\xCon_t) \gCas_{\theta,t}(\xCas_t). \label{eq:case_control_potential:default}
\end{align}

\subsection{Change points}

We again assume a Markov structure for the evolution of the latent states $(\xAll_t)_{t \in \naturals}$. Recall that the latent state at each site~$t$ now takes the form
\begin{align}
 \xAll_t = (z_t, \xCon_t, \xCas_t) = (z_t, \dCon_t, \rCon_t, \dCas_t, \rCas_t) \in \calX \coloneqq \{0,1\} \times (\naturals \times [R])^2,
\end{align}
where the interpretation of $\xCon_t= (\dCon_t, \rCon_t)$ and $\xCas_t = (\dCas_t, \rCas_t)$ is exactly as in a standard change-point model. In particular, at each site~$t$, the regime indicators $\rCon_t$ and $\rCas_t$ take values in $[R]$. The binary latent variable $z_t$ governs the dependence structure between the two groups, i.e.\ $z_t = 1$ indicates that the two groups are \emph{merged}, whereas $z_t = 0$ indicates that they are \emph{split}. More precisely, a~priori, the Markov chain $(\xAll_t)_{t \in \naturals}$ has initial distribution 
\begin{align}
 f_{\theta,1}(x_1) = Q_{\theta,1}(z_1) \fCon_{\theta,1}(\xCon_{1}) \bigl[\ind\{z_1 = 1\} \delta_{\xCon_1}(\xCas_1) + \ind\{z_1 = 0\}\fCas_{\theta,1}(\xCas_{1}|\rCon_{1})\bigr],
\end{align}
on $\calX$ and transition kernels
\begin{align}
 \MoveEqLeft f_{\theta,t+1}(\xAll_{t+1}|\xAll_{t}) \fCon_{\theta,t+1}(\xCon_{t+1}|\xCon_{t})\\
 & \coloneqq Q_{\theta,t+1}(z_{t+1}|\xAll_t) \\
 & \quad \times \bigl[\ind\{z_{t+1} = 1\} {\delta_{\xCon_t}(\xCas_t)}\\
 & \qquad + \ind\{(z_t, z_{t+1}) = (0,0) \,\text{and}\, \rCon_{t+1} \neq \rCas_t\} \fCas_{\theta,t+1}(\xCas_{t+1}|\xCas_{t}; \rCon_{t+1}) \\
 & \qquad + \ind\{(z_t, z_{t+1}) = (0,0) \,\text{and}\, \rCon_{t+1} = \rCas_t\} \fCas_{\theta,t+1}'(\xCas_{t+1}|\xCas_{t}; \rCon_{t+1}) \\
 & \qquad + \ind\{(z_t, z_{t+1}) = (1,0) \,\text{and}\, \dCon_{t+1} \neq 1\} \fCas_{\theta,1}(\xCas_{t+1}|\rCon_{t+1})\\
 & \qquad + \ind\{(z_t, z_{t+1}) = (1,0) \,\text{and}\, \dCon_{t+1} = 1\} \fCas_{\theta,t+1}(\xCas_{t+1}|\xCas_{t}; \rCon_{t+1}) \bigr].
\end{align}
Here, we have defined some of the quantities used above as follows.
\begin{itemize}

 \item The distribution $Q_{\theta,1}(z_1)$ and the transition kernel $Q_{\theta,t+1}(z_{t+1}|\xAll_t)$ have support $\{0,1\}$ and govern the distribution of the latent binary variable.
 
 \item The initial distributions for the two groups, $\fCon_{\theta,1}(\xCon_{1}) \coloneqq \delta_1(\dCon_1) \nuCon_\theta(\rCon_1)$ and $\fCas_{\theta,1}(\xCas_{1}|\rCon_1) \coloneqq \delta_1(\dCas_1) \nuCas_\theta(\rCas_1|\rCon_1)$ are specified as in the single-group scenario. The only non-standard aspect here is that the regime of the case group avoids the regime of the control group, i.e.\ $\smash{\nuCas_\theta(\rCas_1|\rCon_1) = 0}$ for $\rCas_1 = \rCon_1$.
 
 \item The transition kernels
 \begin{align}
 \fCon_{\theta,t+1}(\xCon_{t+1}|\xCon_{t}) 
 & \coloneqq \bigl[\rhoCon_{\theta,t+1}(\xCon_{t}) \delta_1(\dCon_{t+1}) \PCon_\theta(\rCon_{t+1}|\rCon_{t})\\
 & \quad\quad + (1 - \rhoCon_{\theta, t+1}(\xCon_{t})) \delta_{\dCon_{t}+1}(\dCon_{t+1})\delta_{{\rCon}_t}(\rCon_{t+1})\bigr],\\
 \fCas_{\theta,t+1}(\xCas_{t+1}|\xCas_{t}; \rCon_{t+1}) 
 & \coloneqq \bigl[\rhoCas_{\theta,t+1}(\xCas_{t}) \delta_1({\dCas}_{t+1}) \PCas_\theta(\rCas_{t+1}|\rCas_{t}; \rCon_{t+1})\\
 & \quad\quad + (1 - \rhoCas_{\theta, t+1}(\xCas_{t})) \delta_{\dCas_{t}+1}(\dCas_{t+1})\delta_{\rCas_t}(\rCas_{t+1})\bigr],\\
 \fCas_{\theta,t+1}'(\xCas_{t+1}|\xCas_{t}; \rCon_{t+1}) 
 & \coloneqq \delta_1(\dCon_{t+1}) \PCas_\theta(\rCas_{t+1}|\rCas_{t}; \rCon_{t+1}),
\end{align}
are again similar to the single-group scenario. The only non-standard aspect is that for the case group, we enforce that its regime avoids the regime of the control group, i.e.\ $\smash{\PCas_\theta(\rCas_{t+1}|\rCas_{t}; \rCon_{t+1}) = 0}$ for $\rCas_{t+1} = \rCon_{t+1}$.
\end{itemize}
The full state transitions described by the kernel $f_{\theta,t+1}(\xAll_{t+1}|\xAll_{t})$ can be clarified by noting first that the marginal dynamics of the control group $\fCon_{\theta,t+1}(\xCon_{t+1}|\xCon_{t})$ coincide with the single-group model. Second, the transition dynamics of the split indicator $Q_{\theta,t+1}(z_{t+1}|\xAll_t)$ depend potentially on the previous states of the case and control group. Third, conditional on the next state of the control group and the next latent indicator variable, the dynamics of the case group can be motivated by looking at the following different configurations.

\begin{enumerate}

 \item\label{enum:config:a} The two groups are merged at the next site: The case states then coincide with the control states.
 
 \item The two groups have been split, remain split and the control group regime does not jump into the previous regime of the case group: The case states then evolve according to the change-point transition kernel $\smash{\fCas_{\theta,t+1}(\xCas_{t+1}|\xCas_{t}; \rCon_{t+1})}$ with change-point probability function $\smash{\rhoCas_{\theta,t+1}}$ and a regime transition matrix $\PCas_\theta$ that avoids the regime of next control group.
 
 \item The two groups have been split, remain split, but the control group regime jumps onto the previous regime of the case group: The case state then automatically changes the regime according to the regime transition matrix $\PCas_\theta$.
 
 \item The groups have previously been merged and become split next with no change point occurring in the control group: The case group then has a change point sampled from the initial distribution $\fCas_{\theta,1}$.
 
\item The groups have previously been merged and become split next with a change point occurring in the control group: The case state then evolves according to the change-point transition kernel $\smash{\fCas_{\theta,t+1}(\xCas_{t+1}|\xCas_{t}; \rCon_{t+1})}$ as in Configuration~\ref{enum:config:a}.

\end{enumerate}

Conditionally on the model parameters $\theta$, the posterior distribution of interest is then given by
\begin{align}
 p(\xAll_{1:T}|\yAll_{1:T}, \theta) \propto \prod_{t=1}^T \gamma_{\theta, t}(\xAll_{t-1}, \xAll_t),
\end{align}
where $\gamma_{\theta,t}(\xAll_{t-1}, \xAll_t) \coloneqq f_{\theta,t}(\xAll_t|\xAll_{t-1}) g_{\theta,t}(\xAll_t)$.

\subsection{Model parameters}

The transition kernel of the control group is modelled as in the single-group case described in Section~\ref{subsec:model_single_case}, i.e.\ $\fCon_{\theta,t}$ corresponds to $f_{\theta,t}$ therein with the model parameters as in Section~\ref{subsec:model_params_single_case}.

Different choices for the transition kernel of the latent variable $z_t$ are possible. As a first example, we consider a shifted geometric distribution for the duration between change points of $z_t$. Suppose that 
\begin{align}
    Q_{\theta,t+1}(z_{t+1}|\xAll_t)
    & \coloneqq 
    \begin{cases}
     q_{z_t+1, z_{t+1}+1}, & \text{if $\dCas_t \wedge \dCon_t \geq \uAll$,}\\
     \delta_{z_t}(z_{t+1}), & \text{otherwise,}
    \end{cases}
\end{align}
where $q_{i,j}$ denotes the element $(i,j)$ of a $2 \times 2$ row-stochastic matrix 
\begin{align}
 \begin{bmatrix}
  1-\probMerge & \probMerge\\
  \probSplit & 1-\probSplit
 \end{bmatrix},
\end{align}
and $\uAll \geq 0$ is the minimum distance between jumps of the chain $(z_t)_{t \geq 1}$. Although a more flexible model parametrisation for the latent dynamics of the case group is possible in principle that could also be estimated online, we just fix the regime transition probability for the case group via
\begin{align}
\PCas_\theta(\rCas_{t+1}|\rCas_{t}; \rCon_{t+1}) 
 \coloneqq \frac{1}{R-2} \ind\{\rCas_{t+1}\in [R] \setminus \{\rCas_t, \rCon_{t+1}\}\},
\end{align}
and similarly assume for the regime transition density of the split moves that
\begin{align}
\nuCas_\theta(\rCas_{t+1}|\rCon_{t+1})
 \coloneqq \frac{1}{R-1} \ind\{\rCas_{t+1}\in [R] \setminus \{\rCon_{t+1}\}\}.
\end{align}
The probability $\rhoCas_{\theta,t+1}(\xCas_{t})$ of a change point in the case group occurring under the scenario $(z_t, z_{t+1}) = (0,0)$ and $\rCon_{t+1} \neq \rCas_t$ is modelled analogously to Section~\ref{subsec:model_single_case}, i.e.~as the hazard function of a probability mass function $\hCas_{\theta, \rCas_{t}}$, with $\hCas_{\theta, \rCas}$ a shifted negative-binomial probability mass function $\hCas_{\theta, r}(\dCas) = \dNegBin(\dCas-\uCas_r;\kappaCas_r,\omegaCas_r)$ as in Section~\ref{subsec:model_params_single_case}.

Since splits are in practice likely to occur only very infrequently, it would be difficult to estimate any of the case-group specific parameters from the data. Therefore, we fix $\kappaCas_r = \kappaCas = 2$ and assume that $\omegaCas_r = \omegaCas \in (0,1)$ and $\uCas_r = \uCas \geq 0$, where $\uCas \geq 0$ and $\omegaCas > 0$, along with the probabilities $\probSplit, \probMerge\in (0,1)$, are specified by the user.

\subsection{Inference}
\label{subsec:two_group_inference}

We now discuss how one can efficiently perform inference in the case--control model using \gls{SMC} methods. However, unlike the single-group model, the structure of the case--control model is no longer amenable to the particle-filter for change-point models from \citet{fearnhead2007online}. However, the state space is still discrete, and this allows us to retain the idea of deterministically `proposing' particles at each time step in such a perfectly stratified way. Thus, the \gls{SMC} algorithm proposed in this section can be viewed as an instance of the \emph{discrete particle filter} from \citet{fearnhead1998sequential, fearnhead2003online}. Briefly, our methodology again comprises three elements (see Appendix~\ref{app:sec:case_control} for details):
\begin{enumerate}
  \item \textbf{Filtering.}
  We use a type of discrete particle filter \citep{fearnhead1998sequential, fearnhead2003online,whiteley2010efficient} to approximate the `filtering' distributions $p(x_t|y_{1:t}, \theta)$ for all sites $t \in [T]$. This does not rely on random proposals but explores the state space systematically for any possibility. In the case--control model, there are at most $I = 2R + R^2$ possibilities of how each state can evolve with each possibility probed in the particle filter. To remain computationally feasible also for many \gls{CpG} sites, the particle system is pruned down by resampling $M$ particle lineages before the next systematic exploration step. This proposal is then combined with the optimal finite-state resampling scheme from \citet{fearnhead1998sequential}. The algorithm has a linear complexity in $M I$. More details are given in Appendix~\ref{app:subsec:case_control:filtering}.
  
  \item \textbf{Smoothing.} Our main interest is again in the posterior distributions of the latent variables, $p(x_{s:t}|y_{1:T}, \theta)$ for all $s, t \in [T]$ with $s \leq t$. We are particularly interested in the posterior distribution of the latent indicator variables $z_t$ as $p(z_{s:t}|y_{1:T}, \theta)$ encodes our knowledge about whether the two groups are likely different between site~$s$ and site~$t$. We use the forward filtering--backward sampling algorithm from \citet{godsill2004monte} to efficiently approximate such quantities. More details are given in Appendix~\ref{app:subsec:case_control:smoothing}.
  
  \item \textbf{Parameter estimation.} The model parameters $\theta$ are again unknown. However, by construction, only parameters associated with the control-group part of the model need to be estimated. The structure of the model then motivates the following two-step procedure:
  \begin{enumerate}
      \item We run the online-parameter estimation scheme for the single-group model (applied only to the data from the control group) for each chromosome.
      \item The estimated model parameters are then used within the filtering and smoothing algorithms mentioned above.
  \end{enumerate}
\end{enumerate}
We stress that the filtering and parameter-estimation algorithms again need to browse through the data only once. Thus, their computational complexity is again linear in the total number of \gls{CpG} sites, $T$. The smoothing algorithm requires a backward pass, but its overall complexity remains linear in $T$.

Note also that we need to run these algorithms only once even if we want to identify and test for multiple methylation patterns as described next, because we can reuse the particles from the smoothing step.

%
%


\section{Hypothesis-based discovery of methylation signatures}
\label{sec:identification}

\subsection{Discovery of CpG-wise methylation patterns}

\glsreset{FDR}
\glsreset{DMP}
\glsreset{DMR}

Identification of \emph{\glspl{DMP}} is a multiple-testing problem. The observed methylation signals are spatially correlated and accounting for such spatial dependencies can allow for more efficient multiple-testing procedures. While multiple-testing approaches designed for the independent setting can be valid to control the \emph{\gls{FDR}} also under dependence \citep{benjamini2001control}, such procedures can be overly conservative. In this article, the dependence structure is described by a parametric hidden Markov model. Such an approach can allow for optimal procedures under ideal assumptions that minimize the false non-discovery rate, i.e.\ maximize some form of multiple-testing power, subject to a constraint on the \gls{FDR} \citep{sun2009large}. Such idealised procedures require knowledge of the model parameters and can be approximated by a data-driven procedure using consistent estimates of the generative parameters and the latent states. Although such guarantees do not necessarily hold for the approach suggested here -- some model parameters in the case--control model are not estimated for instance -- it can still perform competitively compared to different methods for detecting \glspl{DMP}.

We consider multiple testing for features at individual \gls{CpG} positions. Examples can be testing for differential methylation in case--control studies that aim to reject the nulls $z_t=1$ in favor of the non-nulls $z_t=0$; or testing for hypomethylation signals for the case group $\mu_{\rCas_t}<\mu_{\rCon_t}$ versus the null hypotheses $\mu_{\rCas_t}\geq \mu_{\rCon_t}$.
More generally, for some random variable $\vartheta_t \in \{0,1\}$, which is a function of the latent state $x_t$, we aim to separate the null hypothesises $\calH^0=\{ t \in T \colon \vartheta_t=0\}$ from the signals $\calH^1=\{ t \in T \colon \vartheta_t=1\}$ in a compound decision framework by specifying a decision rule $\delta(y) \in \{0,1\}^T$, where $\delta_t(y)=1$ indicates that hypothesis $t$ is rejected and $\delta_t(y)=0$ otherwise. Testing for the two groups being in different methylation regimes then corresponds to $\vartheta_t=\ind\{z_t=0\}$, whilst testing for hypomethylation corresponds to $\vartheta_t=\ind\{\mu_{\rCas_t}<\mu_{\rCon_t}\}$. Similarly, variance-based signals can be defined at a single \gls{CpG} site. For instance, the signals $\vartheta_t=\ind\{\sigma_{\rCon_t}>\sigma_{\rCas_t}\}$ and $\vartheta_t=\ind\{\sigma_{\rCon_t}<\sigma_{\rCas_t}\}$ correspond to increased methylation variance in the control and case group, respectively.

In large-scale testing problems, the \gls{FDR}, see \citet{benjamini1995controlling}, together with the \emph{\gls{FNR}} can be seen as generalizations of Type-I and Type-II errors for single hypothesis testing. Let
\begin{align}
    R(\delta) &=\sum_{t=1}^T a_t \ind\{\delta_t=1\}, \\
    V(\delta) &=\sum_{t=1}^T a_t \ind\{\delta_t=1\} \ind\{\vartheta_t=0\}, \\
    U(\delta) &=\sum_{t=1}^T b_t \ind\{\delta_t=0\} \ind\{\vartheta_t=1\},
\end{align}
be the weighted number of rejections, false positive results and false negative results, respectively, for non-negative weights $a$ and $b$ that indicate the severity of individual false positive and false negative decisions. We define the \emph{\gls{FDP}} and the \emph{\gls{FNP}} as
\begin{align}
  \text{FDP}(\delta)=\frac{V(\delta)}{R(\delta)\vee 1} \quad \text{and} \quad \text{FNP}(\delta)=\frac{U(\delta)}{|\calH^1|_b},
\end{align}
where $|\calH^1|_b=\sum_{t \in \calH^1} b_t$ is the weighted size of all signals. 
Note that the decision rule $\delta(y)$ is random as a function of the observations $y$ whose law is parameterized by $\theta$. The \gls{FDR} and \gls{FNR} are given by 
\begin{align}
  \text{FDR}(\delta)=\E_{\theta}\left[\text{FDP}(\delta(y))\right]  \quad \text{and} \quad \text{FNR}(y)=\E_{\theta}\left[\text{FNP}(\delta(y))\right],
\end{align}
where the expectation is with respect to $y$. We also define the \emph{\gls{mFDR}} and \emph{\gls{mFNR}} by
\begin{align}
    \text{mFDR}(\delta)=\frac{\E_{\theta}\left[V(\delta(y))\right]}{\E_{\theta}\left[{R(\delta)\vee 1}\right]}  \quad \text{and} \quad  \text{mFNR}(\delta)=\frac{\E_{\theta}\left[U(\delta(y))\right]}{\E_{\theta}\left[|\calH^1|_b\right]}.
\end{align} 
These error measures are of a frequentist nature, in contrast to their Bayesian counterparts in Bayesian decision theory \citep{muller2007bayesian}.

We illustrate how one can test any signal on the loci level in a compound decision-theoretic framework subject to some control on the marginal \gls{FDR} by adapting previous work on multiple-testing procedures, see for instance \citet{sun2009large, muller2004optimal, muller2007bayesian, xie2011optimal, cui2015hmmseq, wei2009multiple}.

For ease of exposition, we assume a single chromosome of $T$ sites and methylation counts $y_{1:T}$ with $\hat{\theta}$ being the estimated model parameters. In practice, we would consider a pooled analysis \citep{wei2009multiple} that combines and ranks the local \glspl{FDR} defined below across all chromosomes. Let $\smash{\hat{p}_t=p(\vartheta_t=0|y_{1:T}, \hat{\theta})}$ be an estimate of the posterior probability of the $t$th hypothesis being null, which is also known as the local \gls{FDR} or local index of significance. These can be computed using the backward simulation Algorithm \ref{alg:backward_simulation}. 

We consider first the unweighted case with $a_t=b_t=1$ for any $t \in [T]$. Denote the ordered estimates of $\{\hat{p}_{t,}\}_{t\in [T]}$ by $\hat{p}_{(1)} \leq \ldots \leq \hat{p}_{(T)}$  and write the corresponding null hypotheses as $\calH^0_{(1)}, \ldots, \calH^0_{(T)}$. Let $\hat{Q}_k=\frac{1}{k}\sum_{t=1}^k \hat{p}_{(t)}$ be an approximation of the \gls{mFDR} based on rejecting the $k$ hypotheses $\calH_{(j)}$ for $j \in [k]$. The testing procedure that controls the \gls{mFDR} at level $\alpha$ in the oracle setting consists then of the step-up procedure:
\begin{align}
     \text{ Reject all } \calH_{(j)} \text{ for } j \in [k] \text{ where } k=\max \{s \in [T] \colon \hat{Q}_s \leq \alpha\}.
\end{align}

Informative domain knowledge can be incorporated into the testing procedure by setting the weights $a_{1:T}$ and $b_{1:T}$ for the costs and gains of multiple decisions. Examples of such prior beliefs or external information that can be used for choosing $a_t$ or $b_t$ might be the genomic features of site $t$ or the distance in base pairs of \gls{CpG} site $t$ to its neighboring \gls{CpG} sites.

For instance, one can choose $a_t=1$ and let $b_t$ be a decreasing function of the distance to neighboring \gls{CpG} sites of site $t$ to reflect the belief that the methylation status of neighboring \gls{CpG} sites becomes less informative when they are further apart. For the weighted case, we denote by $\hat{N}_t=a_t(\hat{p}_t - \alpha)$ an estimate of the excessive error rate when $\calH_t$ is rejected and define a ranking of the different hypotheses that differs from the unweighted case. The hypotheses and excessive error rates are ranked in increasing value of the statistic
\begin{align} 
J_t= \frac{a_t (\hat{p}_t - \alpha)}{b_t(1-\hat{p}_t)+ a_t |\hat{p}_t - \alpha|}
\end{align}
and denoted $\calH_{(t)}$ and $\hat{N}_{(t)}$, respectively. The step-wise procedure that controls the \gls{mFDR} for general weights \citep{basu2018weighted} is:
\begin{align}
     \text{ Reject all } \calH_{(j)} \text{ for } j \in [k] \text{ where } k=\max \biggl\{s \in [T] \colon \sum_{t=1}^s \hat{N}_{(t)} \leq 0\biggr\}.
\end{align}

\subsection{Discovery of region-wise methylation patterns}

Instead of testing individual locations, and possibly reporting them in terms of clusters if the locations are next to each other, it is also possible to test hypotheses over regions or clusters. We assume first that such regions of interest are known a~priori, say particular genes or annotated regions based on different genomic features. Testing for a global null hypothesis, i.e.\ against the alternative that at least one \gls{CpG} site in a region is differentially methylated, may not support strong scientific conclusions. Conversely, tests against the alternative that all \gls{CpG} sites of a region are differentially methylated can be difficult to reject, particularly for large regions. A compromise can be to test the partial conjunction hypotheses \citep{benjamini2008screening} that the proportion of differentially methylated sites is below some tolerance level $\gamma$, $\calH_k^0: \pi_k\leq \gamma$ versus $\calH_k^1: \pi_k>\gamma$ simultaneously for any region $R_k$ from the regions of interest $\{R_1, \ldots, R_K\}$ and 
\begin{align}
\pi_k=\frac{\sum_{t=1}^T \ind \{t\in R_k\}\ind\{z_t=0\}}{\sum_{t=1}^T \ind\{t \in R_k\}} \label{eq:partial_conjunction}
\end{align}
measures the proportion of differentially methylated \gls{CpG} sites, that is sites with different latent regime states in the case and control group, relative to the total number of \gls{CpG} sites in the region $k$. 

More generally, assume that $\{R_1, \ldots, R_K\}$ are a series of regions of \gls{CpG} sites. Our interest lies in separating the nulls $\calH^0=\{k \in K \colon \vartheta_k=0\}$ from the non-nulls $\calH^1=\{k \in K \colon \vartheta_k=1\}$ for some random variables $\vartheta_k \in \{0,1\}$. Let $\{w_k\}_k$ be a sequence of positive weights, which can be set, for instance, proportional to the number of \gls{CpG} sites or base pairs of each region. Denote by
\begin{align}
    R(\delta)&=\sum_{k=1}^K w_k\ind\{\delta_k=1\},\\
    V(\delta)&=\sum_{k=1}^K w_k \ind\{\delta_k=1\} \ind\{\vartheta_k=0\},\\
    U(\delta)&=\sum_{k=1}^K w_k \ind\{\delta_k=0\}  \ind\{\vartheta_k=1\},
\end{align}
the weighted number of rejections, false positive results and false negative results, respectively. With these definitions, \gls{FDP}, \gls{FNP}, \gls{FDR}, \gls{FNR}, \gls{mFDR} and \gls{mFNR} are defined as in the position-wise testing setting, where the signals $\calH^1$ are weighted with $w$ instead of $b$ for computing the \gls{FNP}. Large false positive clusters are thus contributing more to the \gls{FDR} while identifying a large true positive cluster yields a larger gain in power. We define also the \emph{weighted \gls{TP}} as
\begin{align}
    \text{TP}(\delta)=\sum_{k=1}^K w_k \ind\{\delta_k=1\} \ind\{\vartheta_k=1\}.
    \end{align}
We follow the procedure from \citet{sun2015false} that aims to minimize the \gls{FNP} on the regional level, which is also referred to as the missed cluster rate while controlling the \gls{FDR}. Similar to testing for signals at the position level, we consider an estimate of the posterior probabilities of the nulls $\hat{p}_k = p(\vartheta_k=0|y, \hat{\theta})$ with their ordered statistics written as $\hat{p}_{(1)}\leq \ldots \leq \hat{p}_{(K)}$ and corresponding hypotheses $\calH_{(k)}^0$ with weights $w_{(k)}$. We use Algorithm \ref{alg:backward_simulation} to estimate these probabilities. The step-up procedure to test for region-wise methylation patterns at a targeted \gls{FDR} level $\alpha$ is then

\begin{align}
     \text{ Reject all } \calH_{(j)} \text{ for } j \in [k] \text{ where } k=\max \left\{j \in [k] \colon \frac{\sum_{k=1}^j w_{(k)} \hat{p}_{(k)}}{\sum_{k=1}^j w_{(k)}} \leq \alpha \right\}.
\end{align}

We have so far assumed that we are given a collection of regions. However, we can use some preliminary \gls{CpG}-wise analysis to construct regions of interest. For example, we can define the start or end of a region whenever the marginal posterior probabilities $p(z_t=0|y, \hat{\theta})$ crosses some threshold value from below, respectively from above. A theoretical analysis of Type I error controls for such a two-step procedure is beyond the scope of this work. Such post-selection properties of change-points detection algorithms have received less attention in previous work, but see \citet{benjamini2019selection, hyun2021post} for post-selection inference approaches for methylation data, respectively copy number variations. Such methods are not readily available to the more flexible change-point model suggested in this work, and we will therefore contend ourselves to illustrate experimentally that such a two-step procedure can still yield acceptable \glspl{FDP} in the simulations conducted below. 

In practical settings, one is often interested in testing different methylation patterns in an online manner, without knowing in advance how many patterns one will end up testing, for instance by considering new regions from different annotations or changing the tolerance thresholds. Such settings thus call for an online testing framework, where one would like to control the error rates of the procedure whenever one decides to stop the discovery process. This online decision-making process can be accommodated by an adaptation of the procedure in \citet{gang2021structure} to the dependent setting, inclusive of a barrier level suggested therein that filters out large local \glspl{FDR} in order to save FDR budget for testing new methylation patterns in the future.



\section{Simulation study}

\subsection{Methylation data generation and hyper-parameters}

In order to assess the performance of our models and different benchmark methods to discover differential methylation signatures, we generate a collection of \num{100} simulated methylation data sets. Each data set consists of \num{400000} \gls{CpG} sites for a maximum of \num{10} samples per group, with each sample having on average \num{10} reads per \gls{CpG} site. We consider a variety of methylation dynamics and regimes so that the data-generating model does not necessarily coincide with the specific model chosen by the user, i.e.\ to allow for model misspecification. This misspecification allows for a more robust assessment compared to the different benchmark models. In particular, we choose the ten different methylation regimes shown in Figure~\ref{fig:regimes_simulation} with regime-parameters provided in Table~\ref{tab:regimes_simulation} in Appendix~\ref{sec:app:simulation_parameters}, where more details about the hyper-parameters are provided, such as how we sample a range of different values for the model parameters $\theta$.

For each data set, we first employ the inference strategy described in Section~\ref{subsec:single_group_inference} for the single-group model based on retaining $M=100$ particles after pruning/resampling to estimate the model parameters. Subsequently, we perform the filtering and smoothing step for the case--control model from Section~\ref{subsec:two_group_inference}, now using $M = 50$ resampled particles. Details about the inference procedure can be found in Appendix~\ref{sec:app:simulation_parameters}.


\subsection{Detection of differentially methylated positions}

\paragraph{Testing for different means of CpG-wise methylation signals.}

We benchmark the performance of our approach for testing different mean methylation signals at individual \gls{CpG} positions. In a multiple-testing framework, this corresponds to separating the signals $\mathcal{H}^1=\{t\in T \colon \vartheta_t=1\}$ from the nulls $\mathcal{H}^0=\{t\in T \colon \vartheta_t=0\}$ for the signal process $\vartheta_t=\ind{\{\mu_{\rCon_t} \neq \mu_{\rCas_t}\}}$. Notice that there is no signal if one group is in the chaotic regime, whilst the other is in the partially methylated regime with a mean value of $0.5$. 

\begin{figure}[htb]
\begin{subfigure}{.5\textwidth}
		\centering
		\includegraphics[width=1.0\linewidth]{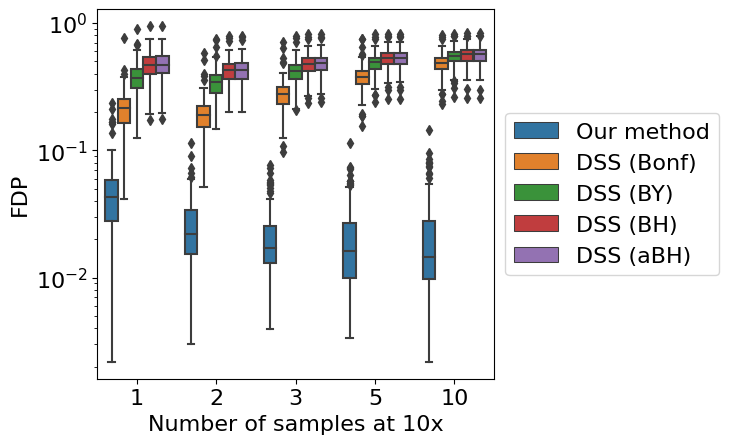}
		\caption{\gls{FDP} for our model and \gls{DSS} with different multiple-testing corrections.}
\end{subfigure}	
\begin{subfigure}{.5\textwidth}
		\centering
        \includegraphics[width=1.0\linewidth]{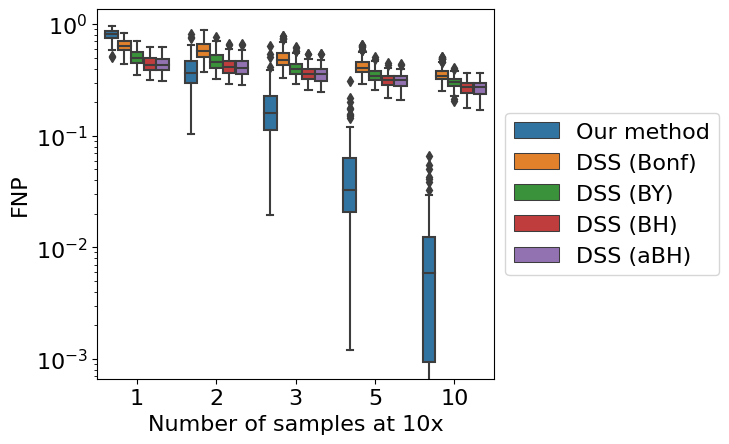}
		\caption{\gls{FNP} for our model and \gls{DSS} with different multiple-testing corrections.}
\end{subfigure}	
\caption{Performance evaluation for testing differences in mean methylation signals over 100 data sets and a target \gls{FDR} of $0.01$. The $p$-value thresholds in \gls{DSS} are based on the different multiple testing corrections: Bonf is a Bonferroni correction, BY is from \citet{benjamini2001control}, BH is from \citet{benjamini1995controlling}, and aBH is from \citet{storey2002direct}. For mathematical definitions, see Appendix \ref{app:p_values}.}
\label{fig:dmp_evaluation_mean_50}
\end{figure}

We, therefore, run \emph{\gls{DSS}} \citep{feng2014bayesian, wu2015detection} on the generated data sets because \gls{DSS} also assumes a beta-binomial model. \gls{DSS} also accounts for some spatial dependencies by applying BSmooth \citep{hansen2012bsmooth} to the methylation state and performing classical multiple testing based on $p$-values. In particular, both our new change-point models and \gls{DSS} allow for variability of the methylation signal for a given mean methylation value. While \citet{wu2015detection} consider a Bonferroni-adjustment, we also consider less conservative procedures, see Appendix \ref{app:p_values} for more precise definitions. The results\footnote{The detected positions by \gls{DSS} depend on a smoothing span parameter. Results for a smaller smoothing span parameter are shown in Figure~\ref{fig:dmp_evaluation_mean_5}.} in Figure~\ref{fig:dmp_evaluation_mean_50} show first that our method yields smaller \glspl{FDP} compared to \gls{DSS}. However, our change-point model leads to realised \gls{FDR}s that slightly exceed the tolerance level of $0.01$, particularly for very small sample sizes. We expect that this is due to the fact that we have allowed for different model misspecifications in the data generation process. Second, our change-point model yields smaller \glspl{FNP} for sample sizes greater than one. Whilst \gls{DSS} achieves lower \glspl{FNP} for a single sample size, this comes at the expense of a very high \gls{FDP}. 

\paragraph{Testing for different means or variances of CpG-wise methylation signals.}

Instead of only testing for differences in the mean methylation level, one can also test for differences in either the mean or the variance of the methylation level at each \gls{CpG} sites. This corresponds to testing for different methylation regimes of the case and control group with signal process $\vartheta_t=\ind{\{\mu_{\rCon_t} \neq \mu_{\rCas_t}\}} \ind{\{ \sigma_{\rCon_t} \neq \sigma_{\rCas_t}\}} = \ind{\{\rCon_t \neq \rCas_t \}}$. The positions detected by \gls{DSS} coincide with those from the previous section. However, as the true signal that we aim to detect changes, we find that the \gls{FDP} for \gls{DSS} decreases for this testing problem, as shown\footnote{Results with a smaller smoothing span parameter for \gls{DSS} are shown in Figure~\ref{fig:dmp_evaluation_regime_5}.} in Figure~\ref{fig:dmp_evaluation_regime_50}. This indicates that \gls{DSS} detects some methylation signals with the same mean and different variances, but attributes this to different mean values. Although \gls{DSS} appears to control the \gls{FDR} in this case, the power is significantly lower compared to the model developed in this work. Our framework also leads to a realised \gls{FDR} that is extremely close to the target level for all sample sizes.

\begin{figure}[htb]
\begin{subfigure}{.5\textwidth}
		\includegraphics[width=1.0\linewidth]{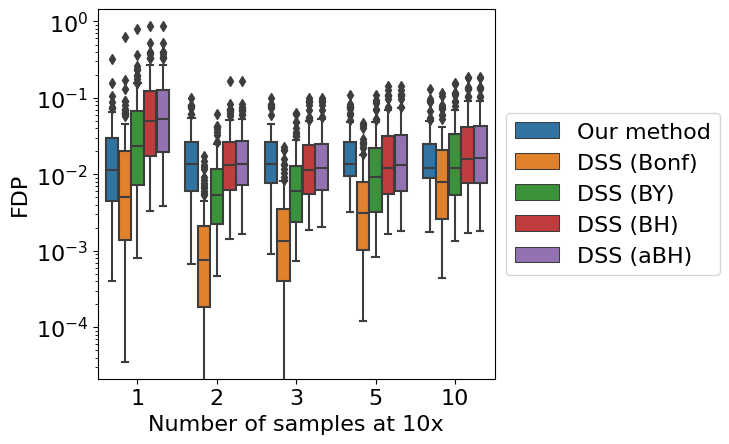}
		\centering
		\caption{\gls{FDP} for our model and \gls{DSS} with different multiple-testing corrections.}
\end{subfigure}	
\begin{subfigure}{.5\textwidth}
		\centering
        \includegraphics[width=1.0\linewidth]{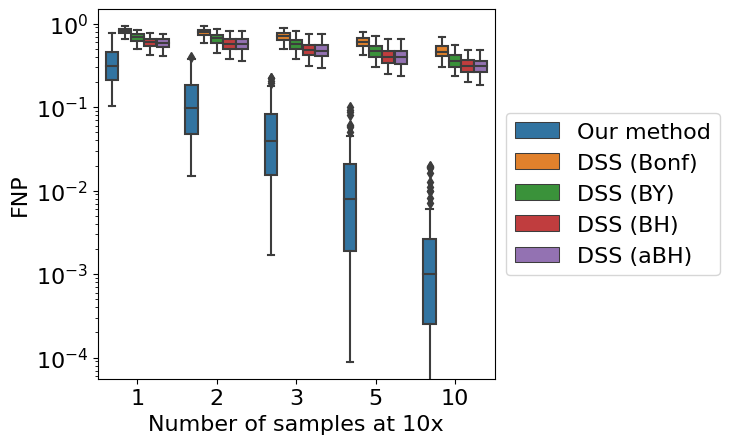}
		\caption{\gls{FNP} for our model and \gls{DSS} with different multiple-testing corrections.}
\end{subfigure}
\caption{Performance evaluation for testing differences in mean or variance of methylation signals over 100 data sets and a target \gls{FDR} of $0.01$.}
\end{figure}
\label{fig:dmp_evaluation_regime_50}

\subsection{Detection of differentially methylated regions}

We consider a two-step procedure for our case--control change-point model that first uses the posterior probability at individual \gls{CpG} sites to construct a possible segmentation of the methylome into regions of interest. Given such regions of interest, we can test for arbitrary methylation signals on such regions. This two-step procedure is similar to that entertained in \citet{korthauer2019detection}, wherein BSmooth \citep{hansen2012bsmooth} is used to smooth the methylation differences, followed by a multiple-testing procedure of such regions of interest. We test for differently strong methylation differences in such regions by varying the percentage of \gls{CpG} sites within each region that must by differentially methylated. More concretely, we consider the partial conjunction tests for the tolerance thresholds $\gamma=0.99$ and $\gamma=0.5$. Although \gls{DSS} can be used to detect differentially methylated regions if a given percentage of \gls{CpG}-based $p$-values exceeds some threshold, this approach lacks any theoretical guarantees for controlling type I errors on the regional level. We emphasize that the weighted sum of the signals $\calH^1$ varies for the constructed regions of interest for the different methods. Indeed, as can be seen in Figure~\ref{fig:drm_positives}, our change-point model yields\footnote{The segmentation of the methylome depends on the threshold used for based on the marginal posterior probabilities of being in different regions. We chose a threshold value of $0.99$ for all tolerance levels $\gamma$.} regions of interest that contain more weighted positives $|\calH^1|_w$. We, therefore, report the \gls{TP} instead of the \gls{FNP} for evaluation purposes. Our change-point model is able to largely control the \gls{FDR} when testing for regions where almost all (i.e.\ \SI{99}{\percent}) of the sites are differentially methylated, as illustrated in Figure~\ref{fig:drm_evaluation_50_regimes_99}, whereas \gls{DSS} and \emph{\gls{DMRSEQ}} \citep{korthauer2019detection} result in very high \glspl{FDP}. At the same time, our change-point model has increased power. When allowing for a lower tolerance level of $\gamma=0.5$, we see from Figure~\ref{fig:drm_evaluation_50_regimes_50} that \gls{DSS} and \gls{DMRSEQ} lead to more moderate \glspl{FDP}. Again, our change-point model achieves a higher power and a lower error rate\footnote{We refer to Figures \ref{fig:drm_evaluation_5_regimes_99} and \ref{fig:drm_evaluation_5_regimes_50} for qualitatively similar results based on using different hyper-parameters for \gls{DSS} and \gls{DMRSEQ}.}.

\begin{figure}[htb]
\begin{subfigure}{.5\textwidth}
		\centering
		\includegraphics[width=1.0\linewidth]{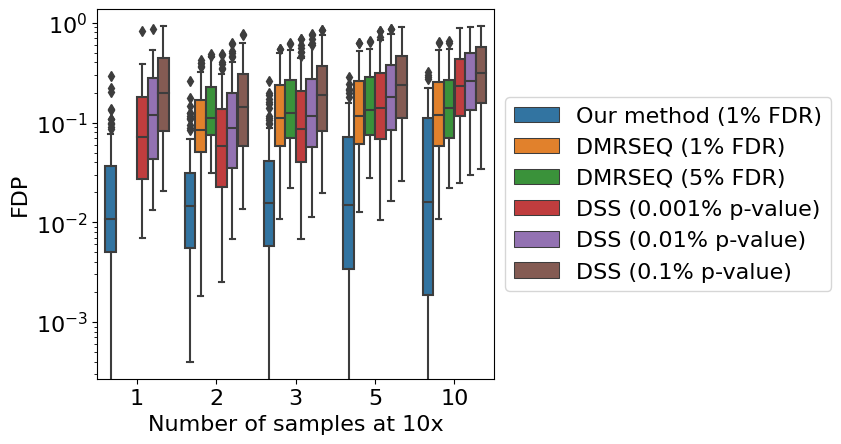}
		\caption{\gls{FDP} for different multiple-testing corrections.}
\end{subfigure}	
\begin{subfigure}{.5\textwidth}
		\centering
        \includegraphics[width=1.0\linewidth]{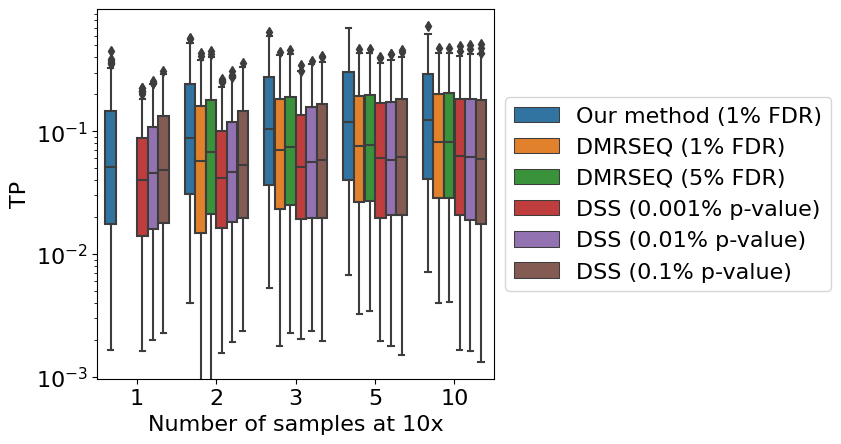}
		\caption{\gls{TP} for different multiple-testing corrections.}
\end{subfigure}
\caption{Performance evaluation for testing region-wise differences in mean or variance of methylation signals for at least \SI{99}{\percent} of \gls{CpG} sites within regions of interest over 100 data sets and a target \gls{FDR} of $0.01$.}
\end{figure}
\label{fig:drm_evaluation_50_regimes_99}

\begin{figure}[htb]
\begin{subfigure}{.5\textwidth}
		\centering
		\includegraphics[width=1.0\linewidth]{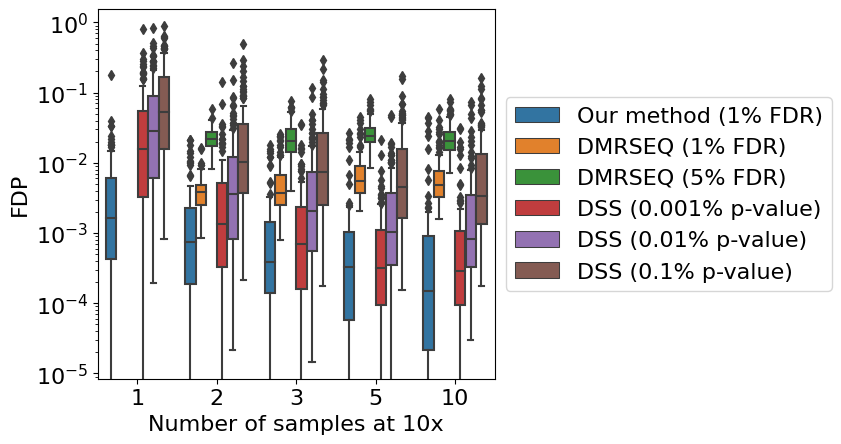}
		\caption{\gls{FDP} for different multiple-testing corrections.}
\end{subfigure}	
\begin{subfigure}{.5\textwidth}
		\centering
        \includegraphics[width=1.0\linewidth]{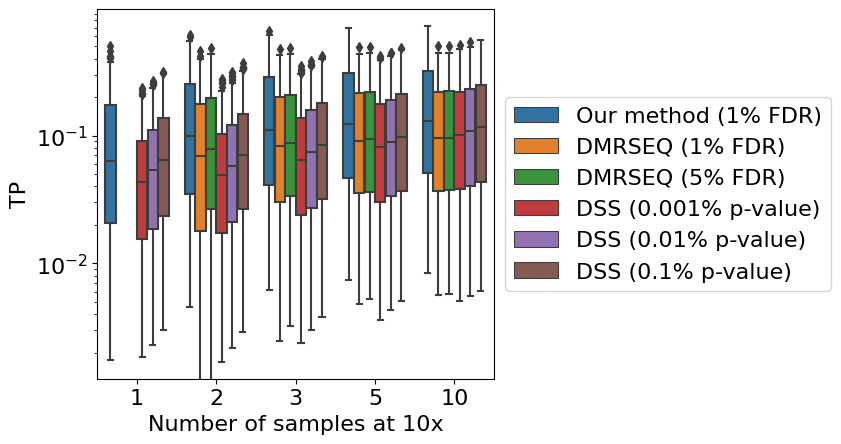}
		\caption{\gls{TP} for different multiple-testing corrections.}
\end{subfigure}
\caption{Performance evaluation for testing region-wise differences in mean or variance of methylation signals for at least \SI{50}{\percent} of \gls{CpG} sites within regions of interest over 100 data sets and a target \gls{FDR} of $0.01$.}
\end{figure}
\label{fig:drm_evaluation_50_regimes_50}

\section{Discussion}

The proposed change-point models for \gls{DNA} methylation as introduced in this article can be adapted in different ways. For instance, we could let the distribution of the sojourn times depend directly on the spatial distance to the most recent \gls{CpG} site, e.g.\ as in \citet{shen2017detect}. The minimum duration requirement between change points could also be easily modified to also include non-\gls{CpG} positions. Furthermore, different transition kernels can be assumed in \gls{CpG}-islands and in regions of low \gls{CpG}-density, respectively.

The suggested standard Bayesian approach is most efficient if the model coincides with the true data-generating mechanism. However, it might not be robust under model misspecification, for instance in the presence of outliers. Alternative belief updates have been suggested using for instance a Tsallis-score \citep{jewson2018principles, knoblauch2018doubly, boustati2020generalised} that downweights observations with a small likelihood which decreases the influence of observations in the tails. The same algorithms can also be used in such a generalised Bayesian setting \citep{bissiri2016general}, but with an adjusted potential function that can be evaluated exactly with not too much additional computational costs if the read depth is not too high. 

It is further possible to adapt the case--control model specifically to twin-pairs data by assuming that the probability parameters $\piCas_{t,s}$ and $\piCon_{t,s}$ coincide whenever $\rCas_{t} = \rCon_{t}$ in order to reduce the observation variance within each twin pair. 
An extension of our models to more general experimental designs will be left for future work, and we expect that a Bayesian framework based on latent variables such as methylation regimes or confounders to be advantageous. 

Forthcoming work will apply the Bayesian change-point models developed in this methodological manuscript to methylation data of type-I diabetes-discordant twins and of a newborn versus an adult.

\section*{Acknowledgements}
The authors acknowledge the use of the UCL Myriad High Performance Computing Facility (Myriad@UCL), and associated support services, in the completion of this work. The authors thank The Alan Turing Institute and the Department of Statistical Science at UCL for supporting MH.
IM is supported by the Biotechnology and Biological Sciences Research Council (grant no. BB/M009513/1).
AB acknowledges support from a Leverhulme Trust Prize.

\setlength{\bibsep}{3pt plus 0.3ex}
\bibliography{./input/bibliography, ./input/jabref}

\appendix

\section{Further details about the single-group model}


\subsection{Parametrisation}
\label{app:subsec:single-group:parametrisation}

\paragraph{Parameters governing the transition matrix.}
To ensure that the matrix $P_\theta$ is stochastic, we parametrise it as a function of model parameters $\theta_{1:((R-1)R)}$ (which are to be estimated from the data) in the following way which also avoids optimisation on the $(R-2)$-simplex, for $r \in [R]$, 
\begin{align}
 P_\theta(r'|r) \coloneqq
 \begin{dcases}
  \dfrac{\exp(\theta_{(R-1)(r-1)+r'})}{\sum_{i=1}^{R-1} \exp(\theta_{(R-1)(r-1)+i})}, & \text{if $r' < r$,}\\
  0, & \text{if $r' = r$,}\\
  \dfrac{\exp(\theta_{(R-1)(r-1)+r'-1})}{\sum_{i=1}^{R-1} \exp(\theta_{(R-1)(r-1)+i})}, & \text{if $r' > r$.}
 \end{dcases}
\end{align}

\paragraph{Parameters governing the sojourn times.}
The regime-specific `success probabilities' in the prior distribution on the sojourn times are parametrised as
\begin{align}
 \omega_r & \coloneqq \mathrm{logit}^{-1}(\theta_{R(R-1)+r}) \in (0,1).
\end{align}

\subsection{Gradient calculations}
\label{app:subsec:single-group:gradient_calculations}

In the remainder of this section, we detail the derivatives which are needed within the online parameter-estimation algorithm.

\paragraph{Parameters governing the transition matrix.}
For any $(r,r') \in [R]^2$ and $j \in [R]$, the gradient for the parameters governing the $j$th row of the transition matrix $P_\theta$ is given by
\begin{align}
	\MoveEqLeft \nabla_{\theta_{(R-1)(j-1):(R-1)j}} \log P_\theta(r'|r)\\
	& = 
	\begin{cases}
		[\iota_{r'}^R-(P_\theta(1|r), \dotsc, P_\theta(R|r) )^\T]_{-r}, & \text{if $j = r$,}\\
		0 \in \reals^{R-1}, & \text{if $j \neq r$,}
	\end{cases}
\end{align}
where $\iota_{r}^R \in \reals^{R}$ denotes a vector filled with $0$'s except for the $r$th element that is equal to $1$; furthermore, the operation $[\ccdot]_{-r}$ removes the $r$th element from a vector, i.e.\ $[x_{1:R}]_{-r} = (x_{1:(r-1)}, x_{(r+1):R})$.

\paragraph{Parameters governing the sojourn times.}
For the conditional change-point probabilities, we have
\begin{align}
	\nabla_\theta \log (1-\rho_{\theta, t+1}(x_t)) 
	& = - \frac{\rho_{\theta, t+1}(x_t)}{1-\rho_{\theta, t+1}(x_t)} \nabla_\theta \log \rho_{\theta, t+1}(x_t),
\end{align}
where
\begin{align}
	\nabla_\theta \log \rho_{\theta, t+1}(x_t)
	& = \nabla_\theta \log \frac{h_{\theta, r_t}(d_{t})}{1 - H_{\theta,r_t}(d_{t}-1)}\\
	& = \nabla_\theta \log h_{\theta, r_t}(d_{t}) + \frac{\nabla_\theta H_{\theta,r_t}(d_{t}-1)}{1 - H_{\theta,r_t}(d_{t}-1)}\\
	& = \nabla_\theta \log h_{\theta, r_t}(d_{t}) + \frac{\sum_{i=1}^{d_t-1} h_{\theta,r_t}(i) \nabla_\theta \log h_{\theta,r_t}(i)}{1 - H_{\theta,r_t}(d_{t}-1)}.
\end{align}
Thus, one only needs to calculate gradients of the form $\nabla_\theta \log h_{\theta,r_t}(i)$. For the regime-specific `success probability' parameters, for any $r' \in [R]$, we have
\begin{align}
	\MoveEqLeft \frac{\partial}{\partial \theta_{R(R-1)+r'}} \log h_{\theta,r}(i)\bigg|_{\omega_{r'} = \logit^{-1}(\theta_{R(R-1)+r'})}\\
	& =  
	\begin{dcases}
		i-u_r - \omega_r (i-u_r +\kappa_r), & \text{$r'=r$ and $i\geq u_r$,}\\
		0, & \text{otherwise.}
	\end{dcases}
\end{align}

\section{Inference for the single-group model}
\label{app:sec:single_group}

\glsunset{IID}

\subsection{Filtering}
\label{app:subsec:single_group:filtering}

The filtering algorithm is known as the \emph{particle filter for change-point models} \citep{fearnhead2007online, yildirim2013online} and is outlined in Algorithm~\ref{alg:model_without_control_group}, where we use the notation $x_t^n = (d_t^n, r_t^n)$ for the $n$th particle at time~$t$. We recall here the expression $\gamma_{\theta,t}(x_{t-1},x_t) = f_{\theta,t}(x_t|x_{t-1}) g_{\theta,t}(x_t)$; $\smash{W_t^n \coloneqq w_t^n / \sum_{m=1}^{N_t} w_t^m}$ denotes the $n$th self-normalised particle weight at time~$t$, for any $n \in [N_t]$, where $N_t$ is the total number of particles employed at time~$t$ (which is not constant over the first few time steps). We remark that our presentation slightly differs from that in \citet{yildirim2013online} where the algorithm targets the one-step-ahead predictive distributions, whereas the algorithm here targets the filtering distributions. In the following, $M \in \naturals$ is a tuning parameter. Both the accuracy of the algorithm but also the computational cost increase with $M$.

\noindent\parbox{\textwidth}{
\begin{flushleft}
  \begin{framedAlgorithm}[Particle filter for the single-group model] ~\label{alg:model_without_control_group}
  \begin{myenumerate}
 \item \label{alg:model_without_control_group:time_1} At time~$t=1$, set $N_1 \coloneqq R$ and:
 \begin{myenumerate}
  \item Set $x_1^n \coloneqq 
  (1, n)$, for $n \in [N_1]$.
  \item Set $\smash{w_1^n \coloneqq \gamma_{1}(x_1^n)}$ and $\smash{W_1^n \coloneqq w_1^n / \sum_{l=1}^{N_1} w_1^l}$, for $n \in [N_1]$.
 \end{myenumerate}
 \item \label{alg:model_without_control_group:time_t} At time~$t$, $t > 1$, set $N_{t-1}' \coloneqq N_{t-1} \wedge M$, $N_t \coloneqq N_{t-1}' + R$.
 \begin{myenumerate}
   
   
  \item If $N_{t-1} > M$, generate $a_{t-1}^{1:M} \coloneqq a^{1:M}$ and $C_{t-1} \coloneqq C$ via the optimal finite-state resampling scheme from Algorithm~\ref{alg:optimal_finite-state_resampling} based on the weights $\smash{W_{t-1}^{1:N_{\smash{t-1}}'}}$.

  
  Else, set $C_{t-1} \coloneqq \infty$ and $a_{t-1}^n \coloneqq n$, for any $n \in [N_{t-1}]$.

  \item For $n \in [N_t]$, set
  \begin{align}
   x_t^n 
   & \coloneqq 
   \begin{cases}
     (d_{t-1}^{a_{t-1}^n} + 1, r_{t-1}^{a_{t-1}^n}), & \text{for $n \leq N_{t-1}'$,}\\
    (1, n-N_{t-1}'), & \text{for $n > N_{t-1}'$,}\\
   \end{cases}\\
   w_t^n
   & \coloneqq 
    \begin{dcases}
     \dfrac{W_{t-1}^{a_{t-1}^n}}{1 \wedge C_{t-1} W_{t-1}^{a_{t-1}^n}} \gamma_{t}(x_{t-1}^{a_{t-1}^n}, x_t^n), & \text{for $n \leq N_{t-1}'$,}\\
     \sum_{m=1}^{N_{t-1}} W_{t-1}^m \gamma_{t}(x_{t-1}^m, x_t^n), & \text{for $n > N_{t-1}'$,}
   \end{dcases}
  \end{align}
  and $\smash{W_t^n \coloneqq w_t^n / \sum_{l=1}^{N_t} w_t^l}$.
 \end{myenumerate}
 \end{myenumerate}
\end{framedAlgorithm}
\end{flushleft}
}


The particle filter relies on the \emph{optimal finite-state resampling} scheme from \citet{fearnhead1998sequential, fearnhead2003online} which we summarise in Algorithm~\ref{alg:optimal_finite-state_resampling}. This method resamples particles (by generating $M$ `ancestor' indices $a^{1:M} \in [N]^M$) based on a set of $N$ normalised weights $\smash{W^{1:N}}$ in such a way that duplicates are avoided.

\noindent\parbox{\textwidth}{
\begin{flushleft}
  \begin{framedAlgorithm}[Optimal finite-state resampling] ~\label{alg:optimal_finite-state_resampling}
  \begin{myenumerate}
   \item Given a set of normalised weights $\smash{W^{1:N}}$, use \citet[][Algorithm~5.2]{fearnhead1998sequential} to find $C > 0$ such that
  \begin{align}
  \sum_{n=1}^{\smash{N}} [1 \wedge C W^n] = M,
  \end{align}
   and let $\smash{n_{1:K}}$ and $\smash{m_{1:I}}$ be the subsequences of $(1, \dotsc, N)$ such that $C_{t-1} W^{n_k} \geq 1$ and $C W^{m_i} < 1$.

  \item For $k \in [K]$, set $a^k \coloneqq n_k$.
  
  \item For $l \in [L]$, set $a^{K+l} \coloneqq b^l$, where $L \coloneqq M - K$ and where $b^{1:L} \in [I]^L$ are generated using systematic resampling based on the weights 
  \begin{align}
   V^i \coloneqq \frac{C}{\smash{L}}W^{m_i}.
  \end{align}
  \item Return $\smash{a^{1:M}}$ and $C$.
 \end{myenumerate}
\end{framedAlgorithm}
\end{flushleft}
}

\subsection{Smoothing}
\label{app:subsec:single_group:smoothing}

Recall that $x_t = (d_t, r_t)$. Our main interest is in computing the marginal posterior probabilities of the regimes, $p(r_{t} = r|y_{1:T},\theta)$, for $t \in [T]$, i.e.\ we want to compute expectations with respect to the marginal smoothing distributions $p(x_t|y_{1:T}, \theta)$, of the following form 
\begin{align}
 \psi_{t|T,r} \coloneqq \E_{x_{t} \sim p(x_t|y_{1:T}, \theta)}[\psi_{t,r}(x_t)],
\end{align}
for functionals $\psi_{t,r}(x_t) \coloneqq \ind\{r_t = r\}$, for all $r \in [R]$ and all $t \in [T]$.
Assuming that the model has sufficient forgetting properties, i.e.~the model is such that $p(x_t|y_{1:T}, \theta) \approx p(x_t|y_{1:(t+\Delta) \wedge T}, \theta)$ for some sufficiently large lag $\Delta \geq 0$, these expectations can be approximated by
\begin{align}
 \psi_{t|(t+\Delta) \wedge T,r} \coloneqq \E_{x_{t} \sim p(x_t|y_{1:t+\Delta \wedge T}, \theta)}[\psi_{t,r}(x_t)],
\end{align}
without introducing too much bias. However, while the bias decreases with $\Delta$, the computational cost of computing (or, at least approximating) these expectations increases with $\Delta$. Thus, a sensible choice of the lag $\Delta$ is crucial. However, manually tuning $\Delta$ is typically  difficult.

Here, we will tune $\Delta$ automatically (and separately for each expectation of interest) via the adaptive-lag smoother from \citet{alenlov2019particle}\footnote{Alternatively, a forward filtering--backward sampling type algorithm \citep{godsill2004monte} could be employed here and would permit the approximation of more general test functions at the cost of including an explicit backward pass. We will use such a strategy for the case--control model below.}. Loosely speaking, this approach exploits that
$\var_{x_{t} \sim p(x_t|y_{1:t+\Delta}, \theta)}[\psi_{t,r}(x_t)] \downarrow 0$ as $\Delta \uparrow \infty$. This motivates approximating $\psi_{t|T,r}$ by $\psi_{t|(t+\Delta_{t,r}(\varepsilon)) \wedge T,r}$, where 
\begin{align}
 \Delta_{t,r}(\varepsilon) \coloneqq \min\bigl\{k \geq 0 \, \big| \, \var_{x_{t} \sim p(x_t|y_{1:t+k \wedge T}, \theta)}[\psi_{t,r}(x_t)] < \varepsilon \bigr\},
\end{align}
for some threshold $\varepsilon > 0$ chosen by the user. Below we summarise the particle-filter approximation based on this idea.

It should be clear that we can use this algorithm to estimate smoothed functionals other  than the posterior probabilities of the regimes. Therefore, in the algorithm given below, we assume that we are interested in a family of $Q$ such test functions at each position, denoted $\{\psi_{t,q}\}_{q \in [Q]}$. In other words, the marginal posterior probabilities of the $R$ regimes can be approximated by taking $Q\coloneqq R$ and $\psi_{t,q}(x_t) \coloneqq \ind\{r_t = q\}$ in Algorithm~\ref{alg:model_without_control_group:smoothing}.

\noindent\parbox{\textwidth}{
\begin{flushleft}
  \begin{framedAlgorithm}[Smoothing the regime indicators] \label{alg:model_without_control_group:smoothing} ~ 
  \begin{myenumerate}
   \item At the end of Step~\ref{alg:model_without_control_group:time_1} of Algorithm~\ref{alg:model_without_control_group}:
 \begin{myenumerate}
  \item Set $\calS \leftarrow \{(1,q) \mid q \in [Q]\}$.
  \item Set $\varPsi_{1|1,q}^n \coloneqq \psi_{1,q}(x_1^n)$, for $n \in [N_1]$ and $q \in [Q]$.
 \end{myenumerate}
  \item At the end of Step~\ref{alg:model_without_control_group:time_t} of Algorithm~\ref{alg:model_without_control_group}:
  \begin{myenumerate}
  \item Set $\calS \leftarrow \calS \cup \{(t,q) \mid q \in [Q]\}$.
  \item For any $(s,q) \in \calS$ and $n \in [N_t]$, set
  \begin{align}
    \!\!\!\!\!\!\!\!\!\!\!\!\!\!\!\!\!\!\!\!\!\!\varPsi_{s|t,q}^n \coloneqq 
    \begin{dcases}
    \psi_{t,q}(x_t^n), &\text{if $s=t$,}\\
    \varPsi_{s|t-1,q}^{a_{t-1}^n}, & \text{if $s< t$ and $n \leq N_{t-1}'$,}\!\!\!\!\!\!\\
    \sum_{m=1}^{\smash{N_{t-1}}} \frac{w_{t-1}^{m} f_{\theta,t}(x_t^n|x_{t-1}^{m})}{\sum_{l=1}^{N_{t-1}}  w_{t-1}^{l}f_{\theta,t}(x_t^n|x_{t-1}^{l})} \varPsi_{s|t-1,q}^{m}, & \text{if $s<t$ and $n > N_{t-1}'$.}\!\!\!\!\!\!
    \end{dcases}
  \end{align}
   \item For any $(s,q) \in \calS$:
   If
   \begin{align}
    \smashoperator{\sum_{n=1}^{N_t}} W_t^n \biggl(\varPsi_{s|t,q}^n - \smashoperator{\sum_{m=1}^{N_t}} W_t^m \varPsi_{s|t,q}^m  \biggr)^{\!\mathrlap{2}} < \varepsilon,
   \end{align}
   set $\calS \leftarrow \calS \setminus \{(s,q)\}$ and return $\smash{\sum_{n=1}^{\smash{N_t}} W_t^n \varPsi_{s|t,q}^n}$ as an estimate of $\psi_{s|T,q}$.
 \end{myenumerate}
\end{myenumerate}
\end{framedAlgorithm}
\end{flushleft}
}

\subsection{Parameter estimation}
\label{app:subsec:single_group:parameter_estimation}

We now describe the additional steps needed to update the parameters $\theta$ via online stochastic gradient-ascent steps. Such algorithms were suggested in \citet{poyiadjis2005maximum, delmoral2010forward, caron2012online}. In addition, \citet{yildirim2013online} proposed an online expectation-maximisation algorithm for the kinds of change-point models discussed here. Write 
\begin{align}
 \phi_{\theta,t}(x_{t-1}, x_t) 
 & \coloneqq \nabla_\theta \log \gamma_{\theta,t}(x_{t-1}, x_t)\\
 & = \nabla_\theta \log f_{\theta,t}(x_t|x_{t-1}) + \nabla_\theta \log g_{\theta,t}(x_t),
\end{align}
with the usual convention that any quantity with site-subscript $0$ is to be ignored from the notation. The required gradient expressions can be found in Appendix~\ref{app:subsec:single_group:parameter_estimation}. With $(\eta_t)_{t \in \naturals}$ denoting some suitable step-size sequence for gradient-ascent algorithms, the parameters can be estimated online using Algorithm \ref{alg:model_without_control_group:gradient-ascent}, which can be extended using adaptive preconditioning such as Adam \citep{kingma2014adam}. Due to the long memory of the change-point model, we found that it can sometimes be advisable to perform the gradient update from Step~\ref{alg:grad_step} only every $\ell$th site (time-step), for some $\ell \in \naturals$.

\noindent\parbox{\textwidth}{
\begin{flushleft}
  \begin{framedAlgorithm}[Parameter estimation] \label{alg:model_without_control_group:gradient-ascent} Choose some initial value $\theta$ at the start of Algorithm~\ref{alg:model_without_control_group}. 
  \begin{myenumerate}
   \item At the end of Step~\ref{alg:model_without_control_group:time_1} of Algorithm~\ref{alg:model_without_control_group}:
 \begin{myenumerate}
  \item Set $\varPhi_1^n \coloneqq \phi_{\theta,1}(x_1^n)$, for $n \in [N_1]$.
  \item Set $\smash{\nabla_1 \coloneqq \sum_{n=1}^{\smash{N_1}} W_1^n \varPhi_1^n}$.
 \end{myenumerate}
  \item At the end of Step~\ref{alg:model_without_control_group:time_t} of Algorithm~\ref{alg:model_without_control_group}:
  \begin{myenumerate}
  \item For $n \in [N_t]$, set
   \begin{align}
    \!\!\!\!\!\!\!\!\!\!\!\!\!\!\!\!\!\!\!\!\!\!\!\!\varPhi_t^n \coloneqq 
    \begin{dcases}
    \varPhi_{t-1}^{a_{t-1}^n} + \phi_{t,\theta}(x_{t-1}^{a_{t-1}^n}, x_t^n), & \text{for $n \leq N_{t-1}'$,}\!\!\!\!\!\!\\
     \smashoperator{\sum_{m=1}^{N_{t-1}}} \frac{w_{t-1}^{m} f_{\theta,t}(x_t^n|x_{t-1}^{m})}{\sum_{l=1}^{N_{t-1}} w_{t-1}^{l}f_{\theta,t}(x_t^n|x_{t-1}^{l})} [\varPhi_{t-1}^{m} + \phi_{t,\theta}(x_{t-1}^{m}, x_t^n)], & \text{for $n > N_{t-1}'$.}\!\!\!\!\!\!
   \end{dcases}
   \end{align}
%
   \item Set $\smash{\nabla_t \coloneqq \sum_{n=1}^{\smash{N_t}} W_t^n \varPhi_t^n}$.
   \item Set $\theta \leftarrow \theta + \eta_t (\nabla_t - \nabla_{t-1})$.
   \label{alg:grad_step}
 \end{myenumerate}
\end{myenumerate}
\end{framedAlgorithm}
\end{flushleft}
}

\section{Inference for the case--control model}
\label{app:sec:case_control}

\subsection{Filtering}
\label{app:subsec:case_control:filtering}


Recall that in the change-point model for the case--control scenario, the state space is given by $\calX \coloneqq \{0,1\} \times (\naturals \times [R])^2$, i.e.\ each state takes the form $x_t = (z_t, \dCon_t, \rCon_t, \dCas_t, \rCas_t) \in \calX$, where $z_t = 1$ indicates that the two groups are merged whereas $z_t = 0$ indicates that they are split. At time~$1$, the model requires that $\dCon_t = \dCas_t = 1$. This implies the following.
\begin{itemize}
 \item At time~$1$, the state $x_1$ can only take one out of (at most) $R^2$ distinct values with positive probability under the model. These values are labelled $\tilde{\chi}_1, \dotsc, \tilde{\chi}_{R^2}$, where
\begin{align}
  \tilde{\chi}_i \coloneqq (\ind\{r = s\}, 1, r, 1, s), \quad \text{for $i = (r-1)R + s$ and $r,s \in [R]$.}
\end{align}

\item At time~$t>1$, conditional on the ancestor particle $x_{t-1}$, the particle $x_t$ can only take one out of (at most) $I \coloneqq 2R + R^2$ distinct values with positive probability under the model. These values are labelled $\chi_1(x_{t-1}), \dotsc, \chi_{2R + R^2}(x_{t-1})$, where
\begin{align}
  \!\!\!\!\!\!\!\!\!\!\!\!\!\!\!\!\!\!\!\!\!\!\!\!\!\!\!\!\!\!\!\!\!\!\!\!\!\!\!\!\chi_i(x_{t-1}) \coloneqq 
   \begin{cases}
    (z_{t-1}, \dCon_{t-1}+1, \rCon_{t-1}, \dCas_{t-1}+1, \rCas_{t-1}), & \text{for $i=1$,}\!\!\!\!\!\!\!\!\!\!\!\!\!\!\!\!\!\!\!\!\\
    (1, \dCon_{t-1}+1, \rCon_{t-1}, \dCon_{t-1}+1, \rCon_{t-1}) \ind\{z_{t-1}=0\} , & \text{for $i=2$,}\!\!\!\!\!\!\!\!\!\!\!\!\!\!\!\!\!\!\!\!\!\!\!\!\!\!\!\!\!\!\!\!\!\!\!\!\!\!\!\!\\
    (0, 1, i-1, \dCas_{t-1} + 1, \rCas_{t-1}), & \text{for $2 < i \leq \rCas_{t-1}+1$,}\!\!\!\!\!\!\!\!\!\!\!\!\!\!\!\!\!\!\!\!\!\!\!\!\!\!\!\!\!\!\!\!\!\!\!\!\!\!\!\!\\
    (0, 1, i, \dCas_{t-1} + 1, \rCas_{t-1}), & \text{for $\rCas_{t-1}+1 < i \leq R+1$,}\!\!\!\!\!\!\!\!\!\!\!\!\!\!\!\!\!\!\!\!\!\!\!\!\!\!\!\!\!\!\!\!\!\!\!\!\!\!\!\!\\
    (0, \dCon_{t-1} + 1, \rCon_{t-1}, 1, i-R), & \text{for $R+1 < i \leq R + \rCon_{t-1}$,}\!\!\!\!\!\!\!\!\!\!\!\!\!\!\!\!\!\!\!\!\!\!\!\!\!\!\!\!\!\!\!\!\!\!\!\!\!\!\!\!\\
    (0, \dCon_{t-1} + 1, \rCon_{t-1}, 1, i-R+1), & \text{for $R + \rCon_{t-1} < i \leq 2R$,}\!\!\!\!\!\!\!\!\!\!\!\!\!\!\!\!\!\!\!\!\!\!\!\!\!\!\!\!\!\!\!\!\!\!\!\!\!\!\!\!\\
    \tilde{\chi}_{i-2R}, & \text{for $2R < i \leq I$.}\!\!\!\!\!\!\!\!\!\!\!\!\!\!\!\!\!\!\!\!\!\!\!\!\!\!\!\!\!\!\!\!\!\!\!\!\!\!\!\!
    \end{cases}
\end{align}
\end{itemize}
where we propose particles with zero values for $i=2$ if $z_{t-1}=1$ to avoid duplicate proposals that have a non-zero transition probability under the prior transition kernel.

In summary, any state sequence up to time~$t$, $x_{1:t}$, can only take one out of (at most) $R^2 I^{t-1}$ distinct values with positive probability under the model. Hence, in principle, filtering and marginal-likelihood computation could be performed analytically in this model by averaging over all $\bo(R^2 I^{t-1})$ state sequences that have positive probability under the model. Unfortunately, the cost of such exact calculations is only feasible if $t$ is very small.

Instead, we follow the idea of the \emph{discrete particle filter} from \citet{fearnhead1998sequential, fearnhead2003online} and include a `pruning' step which ensures that we are only averaging over (at most) $M$ state sequences up to the previous time step at any given time $t$. That is, we set the number of particles at time $t$ to be
\begin{align}
 N_t \coloneqq 
 \begin{cases}
  R^2, & \text{if $t=1$,}\\
  (M \wedge N_{t-1}) I, & \text{if $t > 1$,}
 \end{cases}
\end{align}
where we recall that $I \coloneqq 2R + R^2$ and where $M \in \naturals$ which denotes the maximum number of `resampled' particles and is a tuning parameter which allows us to control the computational cost: We have at most $N_t \leq M I$ particles which means that the algorithm can be implemented in $\bo(MI)$ operations per time step.

The motivation for this specification is that, depending on the value of $M$, we can perform filtering exactly over the first few time steps before we need to start pruning the particle system from $N_{t-1}$ particle lineages down to $M$ particle lineages.

More precisely, if $N_{t-1} \leq M$, we simply extend each of the $N_{t-1}$ existing particle lineages in all $I$ possible directions. If $N_{t-1} > M$, we need to prune the number of particle lineages from $N_{t-1}$ down to $M$ before again extending each existing particle lineage in all $I$ possible directions. If $t \geq 1$ is such that $N_{t-1} \leq M$, then the algorithm reduces to an analytic evaluation of the filtering distributions. Approximations are only induced once $N_{t-1} > M$ because then the total number of possible distinct state sequences up to time $t$ is so large that can no longer keep track of all of them and need to start pruning the number of particle trajectories from $N_{t-1}$ down to $M$.


\noindent\parbox{\textwidth}{
	\begin{flushleft}
		\begin{framedAlgorithm}[Particle filter for the case--control model] ~
		\label{alg:model_with_control_group_adaptive}
			\begin{myenumerate}
				\item  At time~$t=1$, set $N_1 \coloneqq R^2$ and:
				\begin{myenumerate}
					\item For $n \in [N_1]$, set $x_1^n \coloneqq \tilde{\chi}_n$.
					\item For $n \in [N_1]$, set $\smash{w_1^n \coloneqq \gamma_1(x_1^n)}$ and $\smash{W_1^n \coloneqq w_1^n / \sum_{l=1}^{N_1} w_1^l}$, for $n \in [N_1]$.
				\end{myenumerate}
				\item At time~$t$, $t > 1$, set $N'_{t-1} \coloneqq \#\{n \in [N_{t-1}] \mid W_{t-1}^n >0\}$ as well as $N_t \coloneqq (N_{t-1}' \wedge M) I$ and:
				\begin{myenumerate}
				\item Remove particles with zero weights by letting $\smash{n_{1:{N'_{\smash{t-1}}}}}$ be the subsequence of $\smash{(1, \dotsc, N_{t-1})}$ such that $W_{t-1}^{n_{\smash{k}}} > 0$ and writing
				\begin{align}
				    W_{t-1}^{1:N_{\smash{t-1}}'} \leftarrow W_{t-1}^{n_{1:N_{\smash{t-1}}'}}, \quad \text{and} \quad 
				    x_{t-1}^{1:N_{\smash{t-1}}'} \leftarrow x_{t-1}^{n_{1:N_{\smash{t-1}}'}}.
				\end{align}
				\item If $N_{t-1}' \leq M$, for $n \in [N_t]$, set
					\begin{align}
					  a_{t-1}^n &\coloneqq \lceil n / I \rceil,\\
					  x_t^n     &\coloneqq \smash{\chi_{((n-1) \Mod I) + 1}(x_{t-1}^{a_{t-1}^{n}})},\\
					w_t^n &\coloneqq \smash{w_{t-1}^{a_{t-1}^n} \gamma_{t}(x_{t-1}^{a_{t-1}^n}, x_t^n)}.
					\end{align}
				  
           Else, i.e.\ if $N'_{t-1} > M$,
          \begin{myenumerate}
					\item generate $a^{1:M}$ and $C_{t-1} \coloneqq C$ via the optimal finite-state resampling scheme from Algorithm~\ref{alg:optimal_finite-state_resampling} based on the weights $\smash{W_{t-1}^{1:N_{\smash{t-1}}'}}$;

				    \item for $n \in [N_t]$, set
					\begin{align}
					  a_{t-1}^n &\coloneqq \smash{a^{\lceil n / I \rceil}},\\
					  x_t^n     &\coloneqq \smash{\chi_{((n-1) \Mod I) + 1}(x_{t-1}^{a_{t-1}^{n}})},\\
					w_t^n &\coloneqq 
					\frac{w_{t-1}^{a_{t-1}^n} \gamma_{t}(x_{t-1}^{a_{t-1}^n}, x_t^n)}{1 \wedge C_{t-1} W_{t-1}^{a_{t-1}^n}}. 
					\end{align}
				\end{myenumerate}
				\item Set $\smash{W_t^n \coloneqq w_t^n / \sum_{l=1}^{N_t} w_t^l}$.
				\end{myenumerate}
			\end{myenumerate}
		\end{framedAlgorithm}
	\end{flushleft}
}

\subsection{Smoothing}
\label{app:subsec:case_control:smoothing}

We are often interested in making inferences on the latent states in entire genomic regions, say from \gls{CpG} Site~$s$ to $t$, or just at a single site with $s=t$. For a given test function $\psi_{s:t} \colon \calX^{t-s} \to \reals$, we, therefore, aim to compute the expectation with respect to the joint distribution of all latent states in the region given the observations at sites $1$ to $T$ of the whole chromosome, that is $\psi_{s:t|T} \coloneqq \E[\psi_{s:t}(x_{s:t})|y_{1:T},\theta] = \E_{x_{s:t} \sim p(x_{s:t}|y_{1:T}, \theta)}[\psi_{s:t}(x_{s:t})]$. 

In the case--control scenario, it is of particular interest to approximate expectations for test functions $\psi_{s:t}(x_{s:t}) = \ind\{z_s = 1, \dotsc, z_t = 1\}$ under the posterior.  Indeed, for this test function,  $\psi_{s:t|T} = p(z_s = 1, \dotsc, z_t = 1|y_{1:T}, \theta)$, i.e.\ $\psi_{s:t|T}$ the probability of the two groups being in the same regime at \gls{CpG} positions $s, s+1, \dotsc, t$. However, other test functions may be of interest, e.g.\ we may want to take $\psi_{t}(x_t) = \ind\{\rCon_t = r\}$ as in the single-group scenario.

Such expectations cannot be calculated analytically. Instead, we will approximate them using the forward filtering--backward sampling algorithm \citep{godsill2004monte} which is outlined in Algorithm~\ref{alg:backward_simulation}. Given the output (i.e.\ given the particles and weights) generated by the particle filter, Algorithm~\ref{alg:backward_simulation} generates $K$ state sequences $\smash{\hat{x}_{1:T}^1, \dotsc, \hat{x}_{1:T}^K}$ approximately from the posterior distribution $p(x_{1:T}|y_{1:T}, \theta)$. These trajectories can be simulated, potentially in parallel; the computational complexity is $\mathcal{O}(TNK)$, where $N \coloneqq MI$ is the maximum possible number of particles. While $K$ can be chosen arbitrarily both the accuracy of the approximations but also the computational cost grows with $K$; a common guideline is to take $K < N$ \citep{lindsten2013backward}.

\noindent\parbox{\textwidth}{
\begin{flushleft}
  \begin{framedAlgorithm}[Smoothing via backward simulation] \label{alg:backward_simulation} ~
  \begin{myenumerate}
    \item Repeat the following (in parallel) for all $k \in [K]$:
      For $t = T, \dotsc, 1$, 
      \begin{myenumerate}
       \item sample $k_t = l \in [N_t]$ with probability proportional to
            \begin{align}
                \begin{cases}
                    W_T^l, & \text{if $t = T$,}\\
                    \smash{w_{t}^l f_{\theta,t+1}(\hat{x}_{t+1}^{k_{t+1}}|x_t^l)}, & \text{otherwise;}
                \end{cases}
            \end{align}
        \item set $\smash{\hat{x}_t^k \coloneqq x_t^{k_t}}$.
    \end{myenumerate}

   \item For any test function $\psi_{s:t}$, approximate $\psi_{s:t|T}$ by $\frac{1}{K}\sum_{k=1}^K \psi_{s:t}(\hat{x}^k_{s:t})$.
  \end{myenumerate}
  \end{framedAlgorithm}
\end{flushleft}
}

\section{Supplementary material for the simulation studies}

\subsection{Hyper-parameters for data-generation and model inference}
\label{sec:app:simulation_parameters}

\paragraph{Data generation.}
The model parameters $\theta$ used for each data set are sampled as follows: 
We sample $u_{r}\sim \dUnif([3,6])$, $\log \probMerge \sim \dUnif([\log(0.01), \log(0.5)])$, $\log \probSplit \sim \dUnif([\log(0.001), \log(0.05)])$, $\kappa \sim \dUnif([1,3])$, $\omegaCas \sim \dUnif([0.6, 0.99])$ and $\omega_r=0.8$. We draw the non-zeros entries of the columns of the regime transition matrix $P$ from a Dirichlet distribution with concentration parameter $\frac{2}{3}$. Additionally, to construct a data generation mechanism that has the same distribution in both groups, we switch the role of the case and control group for every second simulation.

\paragraph{Parameter estimation for the single-group model.}
For any $r \in [R]$, we fixed a minimum sojourn time of $u_r=3$. We performed a gradient update step after every $\ell=200$ sites with an adaptive learning rate \citep{kingma2014adam} based on a step size of $0.01$ which decays exponentially every $\ell$th site with the rate $0.1$. We browsed through the data twice when estimating the parameters. 

We fixed $\probSplit=0.01$, $\probMerge=0.1$, $\omegaCas=0.8$ and $\uCas=3$. Given the model parameters, we performed the particle filtering and backward sampling algorithm independently over \num{10} different runs. For each run, we used $M=50$ resampled particles and $K=25$ backward particles, yielding an approximation of $p(x_{1:T}|y_{1:T}, \theta)$ based on $250$ particles in total.



\begin{table}[htb!]
 \centering
 \caption{Regime parameters for different data sets.}
 \begin{tabular}{@{}SSSSSSSSSSSr@{}} 
  \toprule
  {$\mu_1$} & {$\sigma_1$} & {$\mu_2$} & {$\sigma_2$} & {$\mu_3$} & {$\sigma_3$} & {$\mu_4$} & {$\sigma_4$} & {$\mu_5$} & {$\sigma_5$} & {$\mu_6$} & {$\sigma_6$} \\\midrule
0.95 & 0.08 & 0.15 & 0.08 & 0.05 & 0.08 & 0.85 & 0.08 & 0.5 & 0.08 & 0.5 & $1/\sqrt{12}$ \\
0.95 & 0.1  & 0.15 & 0.04 & 0.05 & 0.1  & 0.85 & 0.04 & 0.5 & 0.1  & 0.5 & $1/\sqrt{12}$ \\
0.95 & 0.1  & 0.2  & 0.08 & 0.05 & 0.1  & 0.8  & 0.08 & 0.5 & 0.08 & 0.5 & $1/\sqrt{12}$ \\
0.95 & 0.05 & 0.15 & 0.05 & 0.05 & 0.05 & 0.85 & 0.05 & 0.5 & 0.08 & 0.5 & $1/\sqrt{9}$  \\
0.95 & 0.05 & 0.2  & 0.1  & 0.05 & 0.05 & 0.8  & 0.1  & 0.5 & 0.08 & 0.5 & $1/\sqrt{9}$  \\
0.95 & 0.05 & 0.2  & 0.05 & 0.05 & 0.05 & 0.8  & 0.05 & 0.5 & 0.1  & 0.5 & $1/\sqrt{9}$  \\
0.95 & 0.05 & 0.2  & 0.05 & 0.05 & 0.05 & 0.8  & 0.05 & 0.5 & 0.05 & 0.5 & $1/\sqrt{12}$ \\
0.95 & 0.05 & 0.2  & 0.1  & 0.05 & 0.05 & 0.8  & 0.1  & 0.5 & 0.1  & 0.5 & $1/\sqrt{12}$ \\
0.95 & 0.1  & 0.25 & 0.1  & 0.05 & 0.1  & 0.75 & 0.1  & 0.5 & 0.05 & 0.5 & $1/\sqrt{12}$ \\
0.95 & 0.05 & 0.2  & 0.1  & 0.05 & 0.05 & 0.8  & 0.1  & 0.5 & 0.1  & {NA}  & {NA}  \\
\bottomrule
 \end{tabular}
  \label{tab:regimes_simulation}
\end{table}

\begin{figure}[htb!]
\begin{subfigure}{\textwidth}
		\centering
		\includegraphics[width=1.0\linewidth]{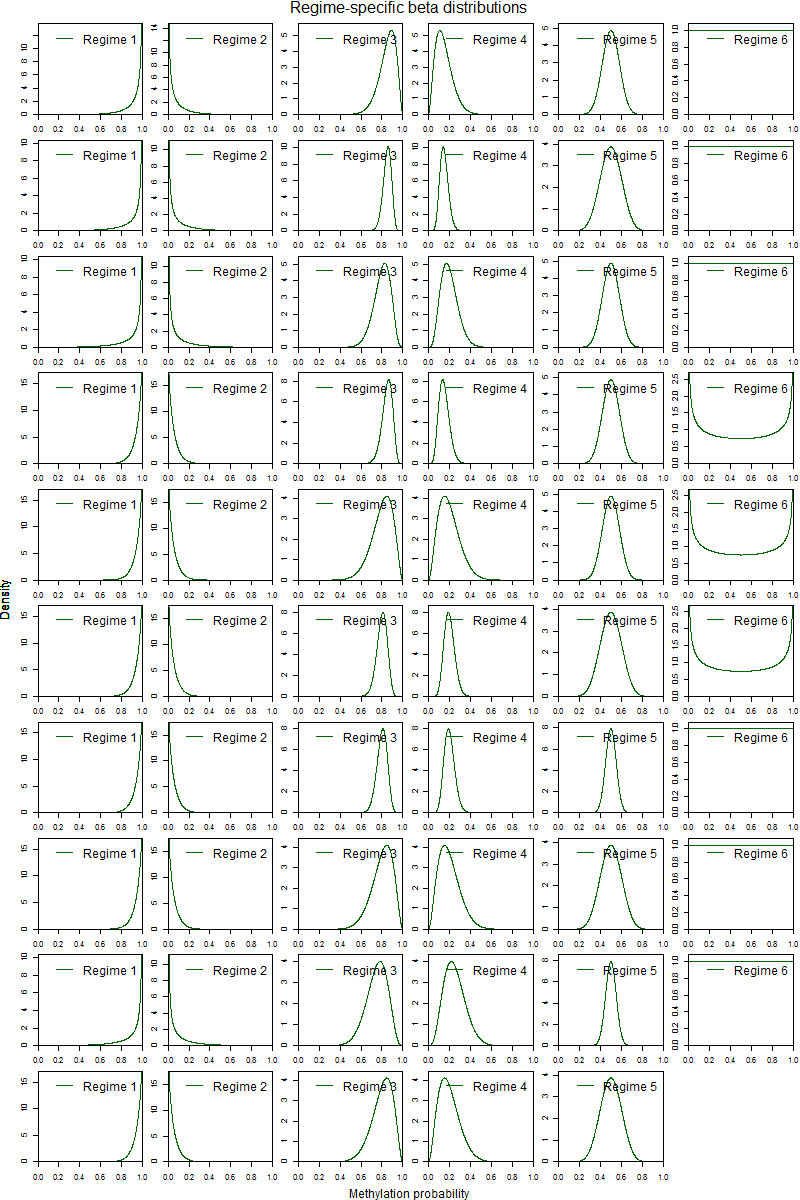}
\end{subfigure}	
\caption{Beta densities for the different regime configurations from Table \ref{tab:regimes_simulation}.}
\label{fig:regimes_simulation}
\end{figure}

\subsection{Multiple-testing corrections and $p$-values}
\label{app:p_values}
We consider the multiple-testing problem of separating the null hypothesises $\calH^0=\{ t \in T \colon \vartheta_t=0\}$ from the signals $\calH^1=\{ t \in T \colon \vartheta_t=1\}$. Suppose that $P_t$ is a $p$-value associated to hypothesis $t$. Let also $P_{(1)}\leq P_{(2)} \ldots \leq P_{(T)}$ be the ordered $p$-values. 
Standard multiple-testing procedures that target an \gls{FDR} of $\alpha$ based on $p$-values reject hypothesis $k$ if $P_k\leq\tau$ for some threshold $\tau>0$. Popular choices are
\begin{enumerate}
    \item Bonferroni procedure (Bonf) that controls even the false-wise error rate at level $\alpha$ for any dependence structure: 
    $$\tau = \max \{ P_{(k)} \colon P_{(k)} \leq \alpha/T\}.$$
    \item \citet{benjamini1995controlling} procedure (BH):
    $$\tau = \max \{ P_{(k)} \colon P_{(k)} \leq k\alpha/T\}.$$
    \item \citet{benjamini2001control} procedure (BY):
    $$\tau = \max \{ P_{(k)} \colon P_{(k)} \leq k\alpha/(\ell_T T)\},$$
    where $\ell_T=\sum_{t=1}^T \frac{1}{T}$.
    \item adaptive BH \citep{storey2002direct} procedure (aBH):
    $$\tau = \max \{ P_{(k)} \colon P_{(k)} \leq k\alpha/(\hat{\pi}_0 T)\},$$
    where $\hat{\pi}_0=\frac{1+ \sum_{t=1}^T \ind\{p_t>\lambda\}}{T(1-\lambda)}$ for some $\lambda \in (0,1)$ is an estimate of the null probability. We set $\lambda=0.5$.
\end{enumerate}

\subsection{Additional results for identifying differentially methylated positions}

\begin{figure}[htb!]
\begin{subfigure}{.5\textwidth}
		\centering
		\includegraphics[width=1.0\linewidth]{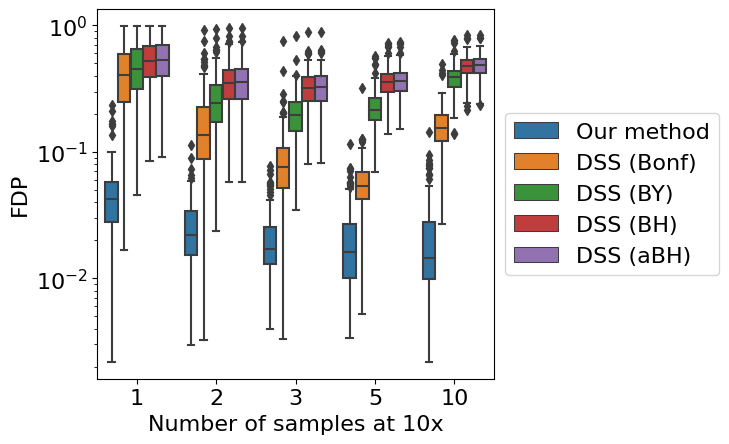}
		\caption{\gls{FDP} for different multiple-testing corrections.}
\end{subfigure}	
\begin{subfigure}{.5\textwidth}
		\centering
        \includegraphics[width=1.0\linewidth]{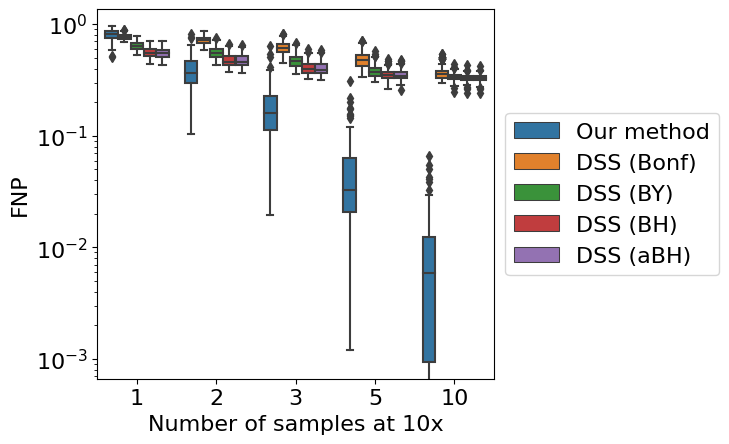}
		\caption{\gls{FNP} for different multiple-testing corrections.}
\end{subfigure}	
\caption{Performance evaluation for testing differences in mean methylation signals over 100 data sets and a target \gls{FDR} of $0.01$. \gls{DSS} uses a shorted smoothing span relative to Figure~\ref{fig:dmp_evaluation_mean_50}.}
\label{fig:dmp_evaluation_mean_5}
\end{figure}

\begin{figure}[htb!]
\begin{subfigure}{.5\textwidth}
		\centering
		\includegraphics[width=1.0\linewidth]{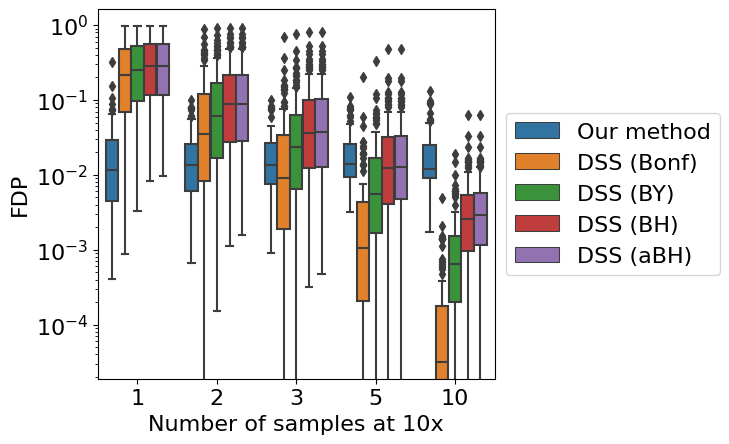}
		\caption{\gls{FDP} for different multiple-testing corrections.}
\end{subfigure}	
\begin{subfigure}{.5\textwidth}
		\centering
        \includegraphics[width=1.0\linewidth]{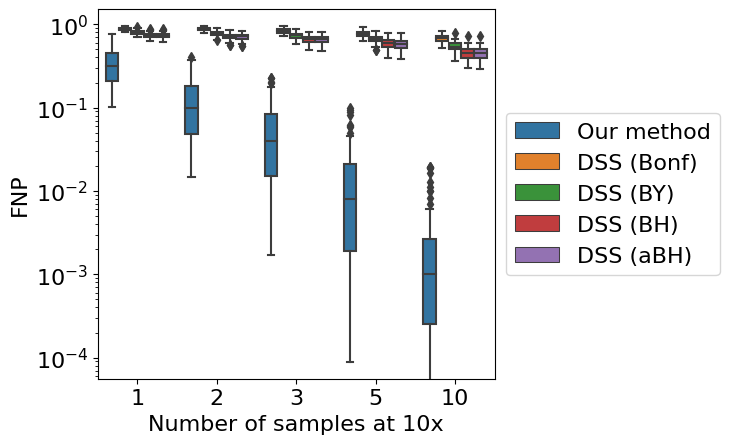}
		\caption{\gls{FNP} for different multiple-testing corrections.}
\end{subfigure}
\caption{Performance evaluation for testing differences in mean or variance of methylation signals over 100 data sets and a target \gls{FDR} of $0.01$. \gls{DSS} uses a shorted smoothing span relative to Figure~\ref{fig:dmp_evaluation_regime_50}}
\label{fig:dmp_evaluation_regime_5}
\end{figure}

\subsection{Additional results for identifying differentially methylated regions}

\begin{figure}[htb!]
\begin{subfigure}{.5\textwidth}
		\centering
		\includegraphics[width=1.0\linewidth]{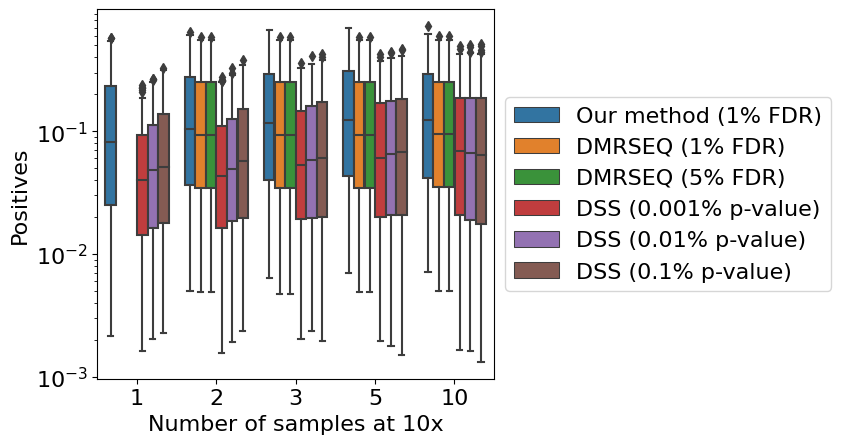}
		\caption{Tolerance level $\gamma=0.99$.}
\end{subfigure}	
\begin{subfigure}{.5\textwidth}
		\centering
        \includegraphics[width=1.0\linewidth]{"simulation_study/ps_0.99_0.99_50_2_False_0.99.png"}
		\caption{Tolerance level $\gamma=0.5$.}
\end{subfigure}
\caption{Positives for the constructed regions of interest for our method, \gls{DSS} and \gls{DMRSEQ} before applying multiple-testing procedures.}
\end{figure}
\label{fig:drm_positives}

\begin{figure}[htb!]
\begin{subfigure}{.5\textwidth}
		\centering
		\includegraphics[width=1.0\linewidth]{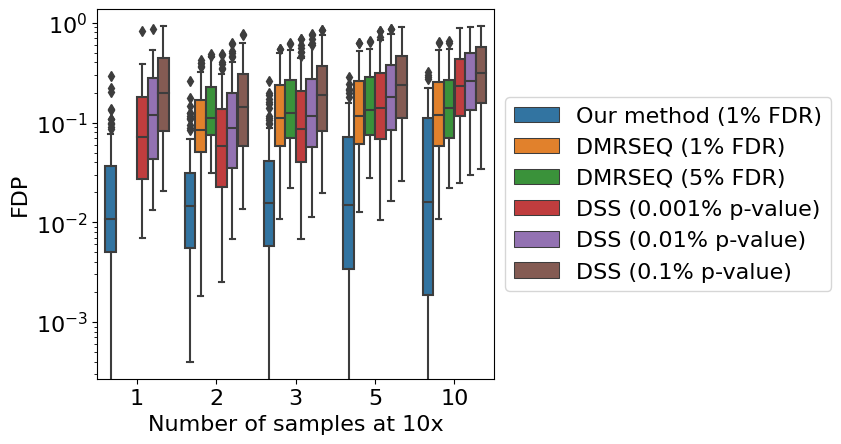}
		\caption{\gls{FDP} for different multiple-testing corrections.}
\end{subfigure}	
\begin{subfigure}{.5\textwidth}
		\centering
        \includegraphics[width=1.0\linewidth]{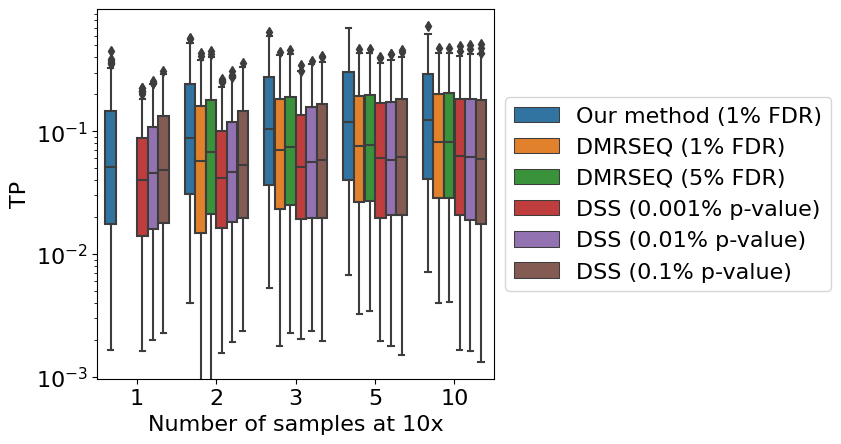}
		\caption{\gls{TP} for different multiple-testing corrections.}
\end{subfigure}
\caption{Performance evaluation for testing region-wise differences in mean or variance of methylation signals for at least \SI{99}{\percent} of \gls{CpG} sites within regions of interest over 100 data sets and a target \gls{FDR} of $0.01$. \gls{DSS} uses a shorted smoothing span, while \gls{DMRSEQ} uses a longer smoothing span relative to Figure~\ref{fig:drm_evaluation_50_regimes_99}.}
\end{figure}
\label{fig:drm_evaluation_5_regimes_99}

\begin{figure}[htb!]
\begin{subfigure}{.5\textwidth}
		\centering
		\includegraphics[width=1.0\linewidth]{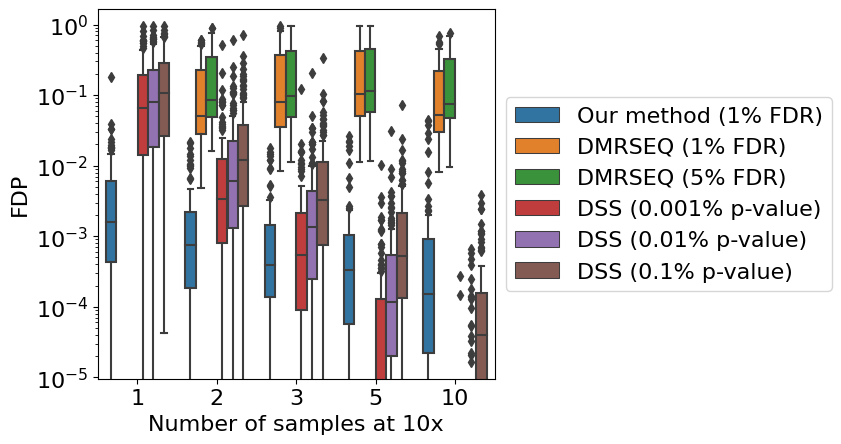}
		\caption{\gls{FDP} for different multiple-testing corrections or q-values.}
\end{subfigure}	
\begin{subfigure}{.5\textwidth}
		\centering
        \includegraphics[width=1.0\linewidth]{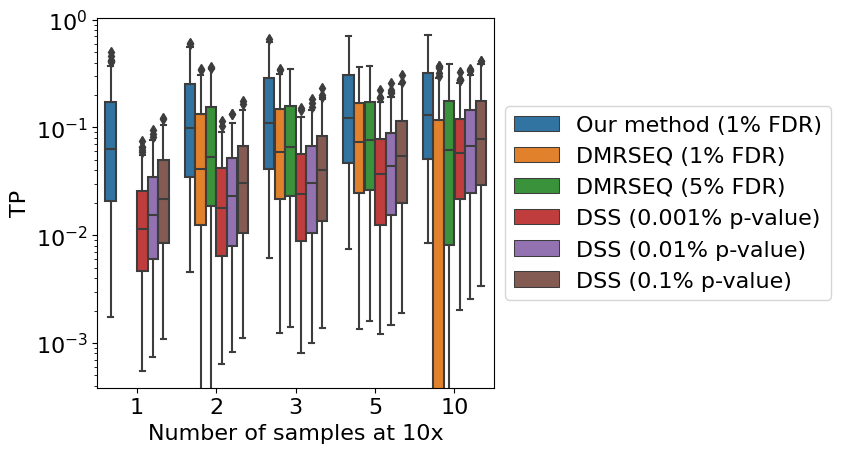}
		\caption{\gls{TP} for different multiple-testing corrections or q-values.}
\end{subfigure}
\caption{Performance evaluation for testing region-wise differences in mean or variance of methylation signals for at least \SI{50}{\percent} of \gls{CpG} sites within regions of interest over 100 data sets and a target \gls{FDR} of $0.01$. \gls{DSS} uses a shorted smoothing span, while \gls{DMRSEQ} uses a longer smoothing span relative to Figure~\ref{fig:drm_evaluation_50_regimes_50}.}
\end{figure}
\label{fig:drm_evaluation_5_regimes_50}

\end{document}